  \documentclass[12pt,onecolumn,twoside]{IEEEtran} 

\usepackage{multirow}
\usepackage{amsfonts}
\usepackage{epsfig}
\usepackage{amsmath}
\usepackage{amssymb}
\usepackage[nolist]{acronym}
\usepackage[english]{babel}
\usepackage{cite}
\usepackage{color}
\usepackage{stfloats}
\usepackage{algorithm}
\usepackage{algorithmic}
\usepackage{subfigure}

\newfloat{ilp}{ht}{aux}
\floatname{ilp}{Integer Linear Programming}

\begin{document}
 \title{Optimal Representations for Adaptive Streaming in Interactive Multi-View Video Systems}

\author{ Laura  Toni,~\IEEEmembership{Member,~IEEE},    and Pascal Frossard,~\IEEEmembership{Senior Member,~IEEE} \thanks{L. Toni and P. Frossard are with  Ecole Polytechnique F\'ed\'erale de Lausanne (EPFL), Signal Processing Laboratory - LTS4, CH-1015 Lausanne, Switzerland. Email: \{laura.toni, pascal.frossard\}@epfl.ch.      }
\thanks{This work was partially funded by the Swiss National Science Foundation (SNSF) under the CHIST- ERA project CONCERT (A Context-Adaptive Content Ecosystem Under Uncertainty), project nr. FNS 20CH21 151569.}
}
\maketitle
\thispagestyle{empty}

\begin{acronym}
\acro{DASH}{dynamic adaptive streaming over HTTP}  
\acro{HAS}{HTTP adaptive streaming}  
\acro{CDN}{content delivery network}  
\acro{CDF}{cumulative density function}
\acro{DP}{dynamic programming}  
\acro{PDF}{probability density function}  
\acro{RTP}{real-time protocol}
\acro{DVC}{distributed video coding}
\acro{QoE}{Quality of Experience}
\acro{VQM}{Video Quality Metric}
\acro{MILP}{mixed integer linear program}
\acro{ILP}{integer linear programming}
\acro{HDTV}{high definition television}
\acro{UGC}{User-Generated Content}
\acro{MPD}{Media Presentation Description}
\acro{IMVS}{Interactive multi-view video streaming}
\acro{DIBR}{depth-image-based rendering}
\acro{VoD}{Video on Demand}
\end{acronym}

\begin{abstract}
Interactive multi-view video streaming (IMVS) services permit to  remotely immerse  within a 3D scene. This is possible by transmitting a set of reference camera views (anchor views), which are used by the clients to freely navigate in the scene and possibly   synthesize additional viewpoints of interest. From a networking perspective, the big challenge in IMVS systems is to deliver to each client the best set of anchor views   that maximizes  the navigation quality, minimizes  the view-switching delay and yet satisfies  the network constraints. Integrating adaptive streaming solutions in free-viewpoint systems offers a promising solution to deploy IMVS in large and heterogeneous  scenarios, as long as the multi-view video representations on the server are properly selected. We therefore propose to       optimize the multi-view data at the server by minimizing   the overall resource requirements, yet offering a good navigation quality to the different users.  We propose a video representation set  optimization for multiview adaptive streaming systems and we show that it is NP-hard. We therefore introduce the concept of multi-view navigation segment that permits to cast the video representation set selection as an integer linear programming problem with a bounded computational complexity. We then show  that the proposed solution reduces the computational  complexity while preserving        optimality   in most of the 3D scenes.      We then provide simulation results for different classes of users and show the gain offered by an optimal multi-view video representation selection compared to recommended representation sets (e.g., Netflix and Apple ones) or  to a baseline representation selection algorithm where the encoding parameters are decided a priori for all the views.
 \end{abstract}

\section{Introduction} 
\label{sec:intro} 
Recent advances in   video technology have opened new research venues     toward novel interactive multi-view services, such as   360-degree videos (e.g., 360 YouTube videos and Google Cardboard\cite{Cardboard}), virtual reality (e.g., Oculus Rift\cite{oculus},   Immerge Lytro camera\cite{Lytro}),  and interactive scene navigation (e.g., free viewpoint interactive TV from BBC\cite{bbc}).
   \ac{IMVS} systems, for example,   endow clients with the ability to choose and display any virtual viewpoint of a 3D scene that has been  originally   captured by images from   a sparse camera arrangement.   This is possible due to the  free-viewpoint technology,    where  a virtual   viewpoint   can be synthesized at decoder via \ac{DIBR}~\cite{tian09} using texture and depth maps of   reference views, namely \emph{anchor views}.  The  quality of the synthesized viewpoints generally increases with both the quality of the anchor views and the similarity between  the anchor views and the   synthesized viewpoint. The quality of the synthesis is thus  improved when   many   high-quality camera views are available, which is however expensive in terms of storage and bandwidth usage.
In real-world environments with limited network resources and heterogenous clients, free viewpoint video streaming   results in  a non-trivial trade-off between network resources and 3D navigation quality.            In such a context, the integration of    adaptive streaming technologies, such as \ac{DASH}, within free-viewpoint systems appears as a promising solution to deploy IMVS in large and heterogeneous scenarios. 
   
\begin{figure}[t]
 \begin{center}
 \includegraphics[width=.65\linewidth,draft=false]{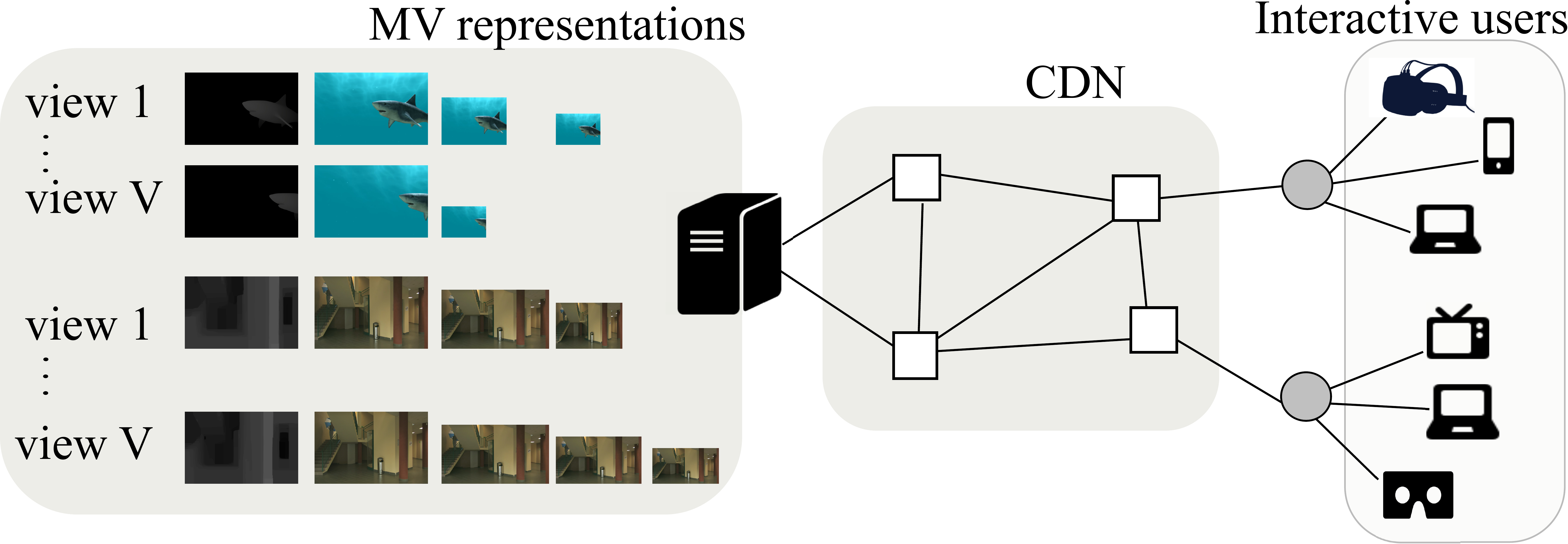}
 \caption{MV adaptive streaming scenario.}
 \label{fig:ConsideredScenario}
 \end{center}
 \end{figure}

 A multi-view adaptive streaming system is depicted in Fig.~\ref{fig:ConsideredScenario}, where several camera views of different video catalogs are stored at the main server of a content provider.  In particular, each camera view consists of a series of texture and depth images. The texture images are pre-encoded in different  \emph{representations}  (i.e., different encoding rates and resolutions) while one version only is pre-encoded for the depth images, due to both their relatively low coding cost and high importance for synthesis\footnote{In this work, we assume undistorted or high-quality depth maps. However, our model can be easily   extended to the case of multiple coding rate also for the depth maps. }. Each representation is then decomposed into temporal  chunks (usually $2s$ long) and then stored at the server. When a client requests a specific multi-view video, it receives a  \ac{MPD}  file from the server, which contains information about the available     representations    for each anchor view.             Given this information, the  client   requests the  best set  of representations for the current chunk based on both its level of interactivity and the available bandwidth.  The best set of representations is defined as the one  that permits to effectively reconstruct a navigation window at the client, namely a range of consecutive virtual views that can potentially be  displayed by the client during the   duration of the video chunk. 
  The requested representations are finally transmitted to the client by the server through a content delivery network (CDN).

 Within this  framework of   multi-view adaptive streaming systems, initial research steps have been made with the study of the    adaptation logic at the client side           to support interactive navigation~\cite{Calagari:C2014, Gao:C2015,Zhao:C15,Su:C14,Su:J2015,Hamza:2014}.   However, the proper selection of  the multi-view video representations at the server has  been    mostly overlooked so far.  Recent results by Netflix \cite{Netflix_blog} and Toni et al. \cite{toni2015optimal} have  however shown   the  importance of properly   adapting the representations  bitrates  in the case of classical video content, in order  to achieve   gains in terms of storage and network costs as well as in terms of video quality   at the client side.   In this paper, we answer to this need and study   the optimal selection of      the multi-view     {representations} that are stored at the server, in such a way that   a good navigation quality  is offered to the   clients when the overall storage and bandwidth resources are limited.  In particular, we argue that,  when the storage capacity at the main server  is constrained, the adaptation of  the multi-view representation set   to  the navigation characteristics in addition to the video content properties    improves the expected  quality experienced by the users during their navigation through the 3D   scene.

In this paper, we present a new framework where we  consider different types of   multi-view video categories, with different video characteristics both in terms of compression and view synthesis properties.
 We then categorize clients into classes,  based on their level of interactivity, their video category of interest  as well as their average downloading bandwidth. 
We then formulate an optimization problem to select the best encoding parameters of the camera views  in order to create multi-view representation sets that permit to maximize the expected satisfaction of the population of interactive users.  The satisfaction function characterizes the navigation quality and depends on both the compression artifacts (driven by the video source rate) and virtual synthesis artifacts (driven by disparity between anchor and virtual views). Our optimization problem is then reformulated as  a novel \ac{ILP} problem with the introduction of the notion of \emph{multi-view navigation segment } that becomes a new optimization variable.   Unlike a mere extension of previous representation set optimization \cite{toni2015optimal}  to MV settings, the proposed ILP problem is tractable in terms of complexity while preserving optimality in most of the common MV scenarios. 
We finally evaluate the benefits of our optimal multi-view representation set selection  in different adaptive streaming scenarios.  
 We show that our optimal set of multi-view representation achieves substantial gain in terms of storage cost and user satisfaction  with respect to recommendation sets.  This means that the   proposed optimization framework reaches an   appropriate balance between user satisfaction and system costs in terms of storage and bandwidth.  This reflects into higher satisfaction of the users with a minimization of  both network and storage resources, which is essential from both a provider and a client perspective.   
 
 In summary, our main contributions are the following:
\begin{itemize}
\item we consider an interactive system with no view switching delay that provides   a navigation window to the users rather than a single view of interest; 
\item we define a satisfaction function for interactive users that captures the quality experienced by users during  their interactive streaming;
\item we formulate a  tractable \ac{ILP} problem based   on the novel concept of multi-view  navigation segments;
  \item we provide extensive simulations results that quantify  the achieved gains  in optimizing the multi-view video  representation set in different adaptive streaming scenarios.
\end{itemize}
To the best of our knowledge, this is the first work that   formulates  a rigorous optimization problem to   design the representation set in adaptive   multi-view  video streaming. It  outlines the importance of adapting the optimal representation set to clients navigation model and content information in interactive systems.  Our optimization is moreover generic and can be integrated in   any  traditional adaptive streaming  infrastructure, like \ac{DASH} or other recent \ac{HAS} systems.   
  Finally, we note that the  proposed optimization is not necessarily meant to be used to make real-time adjustments in DASH systems. It is rather a   framework to derive   optimal encoding solutions for non-live systems or to derive theoretical benchmarks for content providers.

\section{Related Works} 
\label{sec:works} 
In the literature, many works have been published on adaptive streaming on one side and   on interactive multi-view systems on the other side. In the following, we provide only the ones that are  mostly related to our work and we    comment on the recent advances in DASH systems, with a deep focus on $i)$ MV systems,  and  $ii)$ optimization of the representation set for classical videos in \ac{HAS} systems. 

Lately, a deep research has been carried out on MY systems with the goal of improving source coding, streaming strategies, and view synthesis algorithms. 
The main open challenges on  interactive  multiview systems are:  how to efficiently encode distributed sources,   how to efficiently synthesize views,  and  how to efficiently deliver information to users in different applications.   Coding efficiency in MV systems have been studied in the case of  no   communication among cameras,  (e.g., distributed video coding~\cite{DVC}),  or in the case     of joint source coding, (e.g., HEVC for MV \cite{TechChenMul:J16},   graph-based coding and representation \cite{Thomas:j15, motz2016graph}, or coding for interactive users   \cite{Abreu:A15, Dai:J16}).   To improve the process of view synthesis, several works have been focused mainly on improving the \ac{DIBR} technique,  \cite{Gao:J16, Motz15, Thas, Buyssens:2015}. 
 Streaming strategies and network optimization for multiview video, however, is still a relatively unexplored and new research topic.   Optimal streaming strategies or reousrce allocation have been proposed  in the case of  resource-constrained networks  \cite{Toni:A14, cheung11tip2, Toni:C13}, but  not from a  HAS systems perspecitve.   
 A DASH-based stereo 3D video systems is introduced in  \cite{Calagari:C2014}, while a DASH-based   multi-modal 3D video streaming system is proposed  in \cite{Gao:C2015}.   
Within the framework of free viewpoint systems, the works in   \cite{Su:C14,Su:J2015}  design  an architecture of   DASH-based 3D multi-view video streaming services using the  HEVC 3D extension encoder  to prepare the representations at the DASH server. 
   In \cite{Hamza:2014},  a complete architecture for a free viewpoint streaming system based on adaptive HTTP streaming is presented, while  
   a cloud-assisted   DASH-based IMVS  scheme is proposed  in \cite{Zhao:C15}, where view synthesis can be performed either at the    server or at  the client. 
     At the client side, a single viewpoint with adaptive bitrate is requested, while in our work we rather consider a navigation window to enable low-delay view switching.   The concept of navigation window  has been considered also in  \cite{Hamza:C16}, where  authors propose  a complete architecture for HTTP adaptive streaming of free-viewpoint videos. In particular, they design   a   two-step rate adaptation method that takes into consideration the users interaction with the scene (the navigation trajectory of the user is predicted) as well as the special characteristics of multi-view-plus-depth videos and the quality of rendered virtual views. 
   The above  works, however, consider interactive MV  systems     with   pre-encoded   representations  according to classical recommendations, with fixed coding parameters and do not   investigate the impact of these sets in different DASH settings.  Our work is therefore complementary to the above     ones since it studies  adaptive streaming systems from a provider's perspective   and optimizes the multi-view video representations at the server.

In the framework of HTTP adaptive streaming of mono-view videos,  recent works have been carried out on the optimal design of the representation set in DASH systems~\cite{Cong_Thang:JSAC14, Aparicio-Pardo:2015:,Zhang:J13, Essaili:J15,toni2015optimal,Netflix_blog}. 
In~\cite{Cong_Thang:JSAC14},   authors  show how the selection of  representations sets may affect   the behaviors of some adaptation logics in live-streaming DASH systems, motivating  the interest of optimizing the representations set.  A formal optimization of  the representations set however is not provided in the paper. Such an  optimization is proposed in~\cite{Aparicio-Pardo:2015:} for cloud-based  live video streaming applications, in~\cite{Zhang:J13}  for designing optimal caching strategies over the CDN, and in~\cite{toni2015optimal} for designing the optimal representations sets  to be stored at the main server. 
In   \cite{toni2015optimal}, the authors  propose  an \ac{ILP} problem for adapting the optimal  representation  set to video content characteristics and users population, showing the sub-optimality of   the commonly  recommended representations (YouTube, Netflix and Apple) with respect to  the resulting optimal   set.  
Among those, the  work   \cite{toni2015optimal} shares  similarities with  our work in terms of the overall goal, however it cannot be directly extended to free viewpoint adaptive streaming, since the viewpoint synthesis  and the  interactivity of the clients need to be properly   be taken into consideration in the optimization.  

 \section{System Model}
\label{sec:system_model} 
We now describe the system model for the  \ac{VoD} adaptive streaming system both from a server and client side perspective. 

At the \emph{main server},   several multi-view video sequences are stored  and provided  to interactive clients.    For each video content $c\in \mathcal{C}$,   we denote by $\mathcal{V}_c=\{1, 2, \ldots, V_c\}$   the set of    camera views acquiring  the scene over time.    Each view is composed of   both a texture image  and   a depth map. We assume that texture images are actually encoded in different  \emph{representations} (i.e., different encoding rates and spatial resolutions) while  depth maps are   stored only at good quality to avoid inconsistency in view   synthesis. \footnote{While coding parameters for texture and depth images can be jointly optimized, in practice
$i)$ depth maps are usually much more efficiently compressed than texture images;  $ii)$ for a given content, the optimal ratio between texture and depth data remains roughly  the same for any total target bit-rate~\cite{Bosc2013}. Therefore depth maps are usually neglected in MV coding optimizations~\cite{AbreuCF016}. }   Therefore a constant coding rate is added to the coding rate of the texture image,    which represents the overhead of the depth map. However, our problem is general enough and it can be easily extended to the case of multiple coding rates for both texture images and depth maps, as shown at the end of the section.   
 Denoting by $\mathcal{R}$ and   $\mathcal{S}$  the set of possible coding rates and spatial resolutions for each texture image, we define by $\mathcal{L}_c$ the set of all possible representations that can be generated from the encoder. More formally, $\mathcal{L}_c = \{ l :(v,r,s) | v\in \mathcal{V}_c, r\in \mathcal{R}, s\in \mathcal{S} \}$, where the triple $(v,r,s)$ identifies a representation of camera view $v$ encoded at rate $r$ and resolution $s$. Storing all representations in $\mathcal{L}_c$  for all video sequences in  $\mathcal{C}$ might be too expensive. Therefore, only a subset of representations is actually encoded and stored in practice. We denote by  $ \mathcal{T}_c   \subseteq \mathcal{L}_c$ the subset of representations stored at the main server for video content $c$.

\begin{figure}[t]
 \begin{center}
 \includegraphics[width=.75\linewidth,draft=false]{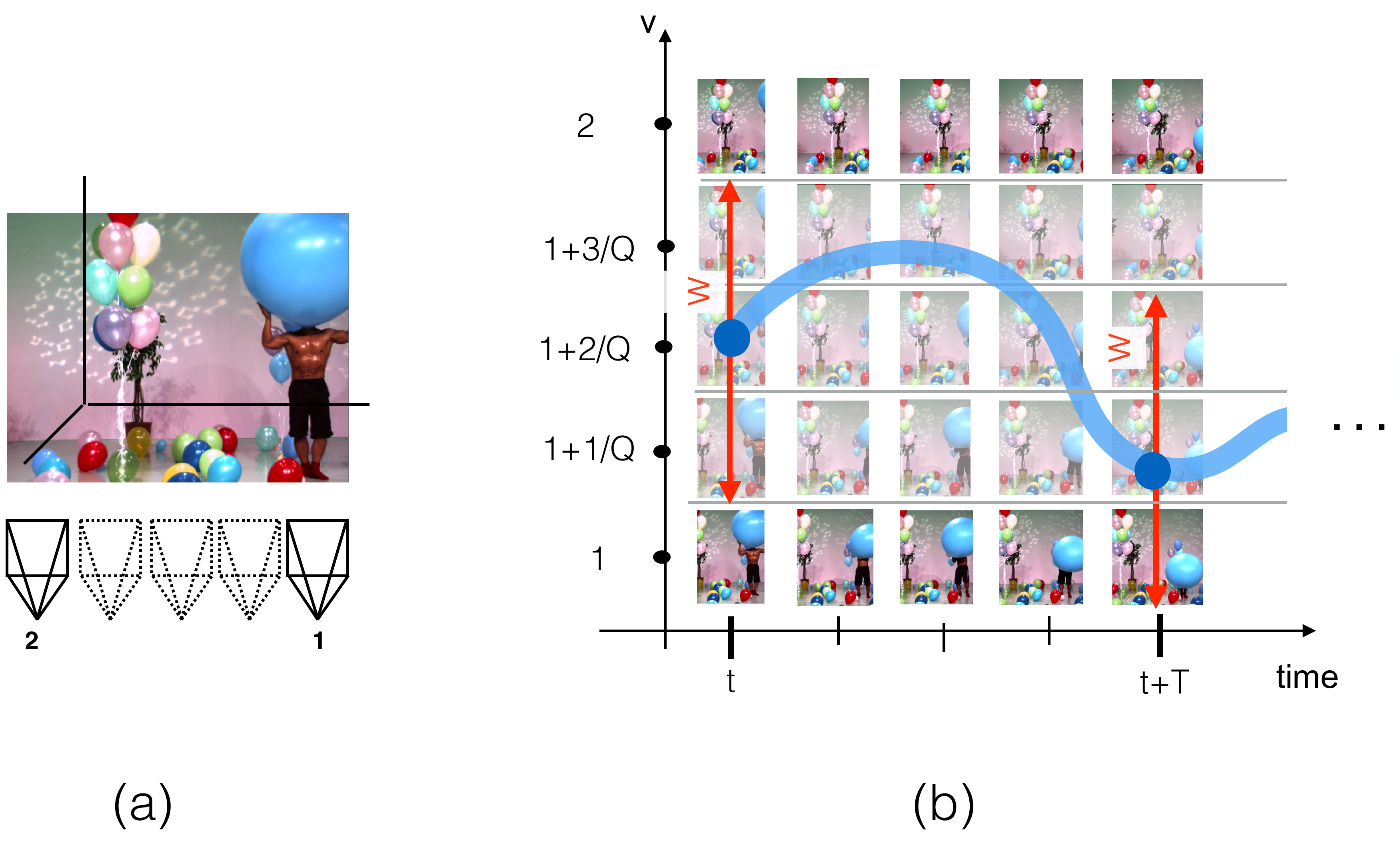}
 \caption{Example of navigation windows at two consecutive downloading opportunities.  (a) Scene at time $t$ acquired by two camera views ($1$ and $2$).  (b) Navigation path  and navigation windows of a user at the downloading time $t$ and $t+T$ in the case of three virtual viewpoints between consecutive camera views (virtual viewpoints are $1+i/Q$, with $i=1,2,3$).  The starting viewpoint at $t$ is $u=1+2/Q$ leading to a navigation window  $w(1+2/Q)=[1+1/Q,1+3/Q]$. At time $t+T$, the starting viewpoint is $u=1+1/Q$ with $w(1+1/Q)=[1,1+2/Q]$.}
 \label{fig:NW}
 \end{center}
 \end{figure}

  Each representation is segmented into temporal chunks and   at the \emph{client side}, each client   sends  a periodic downloading request every $T$ seconds,  which is the duration of a temporal chunk\footnote{We assume to be at regime (not  during the startup or rebuffering phases) when  downloading opportunities are periodic, one downloading request every chunk duration. }. 
  We call starting viewpoint the viewpoint $u$ displayed at each downloading opportunity, with $u\in[1,V_c]$ for user displaying content $c$. Within the following $T$ seconds, the user is free to navigate in the scene. Denoting by $\rho$   the maximum  speed at which a user can navigate to neighboring virtual views,   the user can display a   range of viewpoints  defined as $w(u)=[u - \rho T, u + \rho T]$ with $u$ being the starting viewpoint, as shown in Fig.   \ref{fig:NW}. 
 To ensure  a free scene navigation  with no switching-delay, the client   chooses to download     representations  that   permit to reconstruct all viewpoints in  the navigation window $w(u)$. 
Given a set of dowloaded representations, any virtual viewpoint $u$ can be synthesized using a pair of left and right reference view images $v_L$ and $v_R$ in the downloaded set, with  $v_L < u < v_R$, via a classical DIBR techniques, e.g.,~\cite{Mori200965}.   We denote by $v$ any camera view (thus any possible  anchor view) while $u$ represents any viewpoint (either virtual viewpoint or camera view) that can be displayed  during the navigation.  We also denote by    $\mathcal{U}_c=\{ 1, 1+1/Q, 1+2/Q,  \ldots, {V_c} \}$ the set of all viewpoints that can be displayed for video content $c$.

We then categorize   users into classes,   characterized by the type of  rendering device, the download bandwidth, the navigation characteristics,   as well as the required video content. In particular, a \emph{user of type $i$ }displays the video of interest $c_i\in\mathcal{C}$ on a device with spatial resolution $s_i$ and  a network connection with a maximum downloading rate  $B_i$.  Furthermore, we denote by $q_{i}(u)$   the probability for this user to display the starting viewpoint $u$.    This means that a user of type $i$ will be interested in reconstructing the  navigation window $w(u)$ centred in $u$  with probability $q_{i}(u)$  and will therefore select the best set of representations   with spatial resolution  $s_i$ and   an average downloading rate lower than $ B_i$.  Note that  we assume no   switching  in the spatial resolution of images  at the decoder and we impose a hard  network constraint of $B_i$. Possible client strategies for selecting the best set of representations are \cite{Su:J2015,Hamza:2014,Zhao:C15}. 
Finally, we denote by  $\zeta_{i}$   the portion of clients of type-$i$,  with   $\sum_{i\in \mathcal{I}} \zeta_{i}=1$, where  $\mathcal{I}$ is  the set of all possible types of    clients.

We   remake that  our   model  can be easily extended to  more complex systems. For example, one can have several   quality levels for the depth maps and a representation $l$   defined as  $l:(v, r^\text{t}, r^\text{d}, s)$, with  $r^\text{t}, r^\text{d}$ being the encoding rate for texture and depth images, respectively.  Furthermore, the case of spatial  resolution switching at the decoder can be modeled  by  introducing the resolution switching artifacts in the distortion model.  Finally, our model has a very simple communication channel representation, which abstracts from elements such as sending and receiving buffers. However, even if  the presence of the buffer reflects with higher fidelity  the behavior of realistic DASH clients, it has mainly an impact on the client strategy rather than on the coding strategy at the server side \cite{toni2015optimal}, which is the problem of interest of this paper.

 \section{Optimal MV Representation Set}
\label{sec:ILP} 
We now describe the   optimization of the representation set for a multi-view adaptive streaming. First, we provide the   problem formulation. Then, we introduce the concept of multi-view navigation segment  and finally we cast the representation optimization  as an \acf{ILP} problem  with a tractable complexity solution.

\subsection{Problem Formulation}
Given a set of videos  $\mathcal{C}$ and a set of possible representations $\mathcal{L}=\bigcup_c \mathcal{L}_c$ for each chunk   under consideration, we seek the best subset of representations $\mathcal{T}^*\subseteq \mathcal{L}$ that can be stored at the  server such that the   average navigation quality  offered to  the users is maximized.   We  can write  our \emph{MV representation set optimization} problem as follows  
\begin{align}\label{eq:problem_form}
\mathcal{T}^* : &
\arg \min_{\mathcal{T}} \sum_{i\in \mathcal{I}}  
\left\{ \sum_{u \in  \mathcal{U}_{c_i}}   \zeta_{i}  \, q_{i}(u) D_{c_i}(w(u),  \mathcal{T}, B_i)\right\}
\nonumber \\
&\text{s.t. }      \sum_{l\in \mathcal{T}}  r_{l} \leq |\mathcal{C}| C 
\end{align}
where  $ |\mathcal{ C}| C$ is the global storage constraint  (with $C$ being the mean storage constraint per video)\footnote{The storage constraint takes into account also size of the depth maps. However, for the sake of clarity, we omit this constraint in our formulation. }, and $D_{c}(w,  \mathcal{T}, B)$ is the best distortion experienced by the client displaying video $c$ with the navigation window $w$\footnote{The navigation window $w(u)$ depends on the starting viewpoint $u$. However, for the sake of clarity, in this section we omit the dependency from $u$ when not needed.  }, given that the subset of representations $\mathcal{T}$ is available and that a maximum downloading rate of $B$ is imposed.  More formally, 
\begin{align}\label{eq:mean_dist}
D_c\left(w, \mathcal{T}, B \right)   &= 
 \min_{\substack{\mathcal{O}\subseteq \mathcal{T}:  \\ \sum_{l \in \mathcal{O}} r_{l } \leq B}}
 \left\{ \sum_{u\in {w}} 
 d_{u,c}^{\star}(\mathcal{O})  
   \right\}
\nonumber  \\ 
 &=\min_{\substack{\mathcal{O}\subseteq \mathcal{T}_c:  \\ \sum_{l \in \mathcal{O}} r_{l } \leq B}}
 \left\{ \sum_{u\in {w}} 
  \min_{ \substack{ [l_L, l_R]\in\mathcal{O}: \\ v_L \leq u, v_R\geq u } }  d_{u,c}(l_L, l_R)
  \right\} 
\end{align}
where $\mathcal{O}$ is the set of representations downloaded by the user,  $d_{u,c}^{\star}(\mathcal{O})$ is the  distortion of the viewpoint $u$ of video $c$ synthesized by a pair of anchor views from the ones available in $\mathcal{O}$, and $d_{u,c}(l_L, l_R)$  is the  distortion of the viewpoint $u$ of video $c$ synthesized from the representations $(l_L, l_R)$. 

 The  optimization problem in  \eqref{eq:problem_form} is computationally complex to solve. \footnote{Note that   the above optimization problem is  refined  periodically with a frequency equivalent to the duration of $K$ chunks.   Obviously,  the lower the $K$, the more refined   the representation set.  But each optimization becomes more  complex.}   
It can be proved to be NP-hard actually  by considering   a special case with  only one client    with infinite bandwidth. In this special case, the representation selection problem reduces to a camera  selection problem, which  can be   cast as the known NP-hard set cover (SC) problem \cite{toni2015network}. Solving the special case of the representation selection problem is no easier than solving the SC problem, and therefore it is also NP-hard. The more general   representation selection problem   in \eqref{eq:problem_form} is no easier than the special case, and therefore the problem    in \eqref{eq:problem_form}  is NP-hard. Therefore, in the next subsection, we show how to cast \eqref{eq:problem_form}   in a tractable optimization problem.   With this aim, we first introduce a new optimization variable.

\subsection{MV Navigation Segment}
\begin{figure}[t]
   \begin{center}  
 \includegraphics[width=.45\linewidth,draft=false]{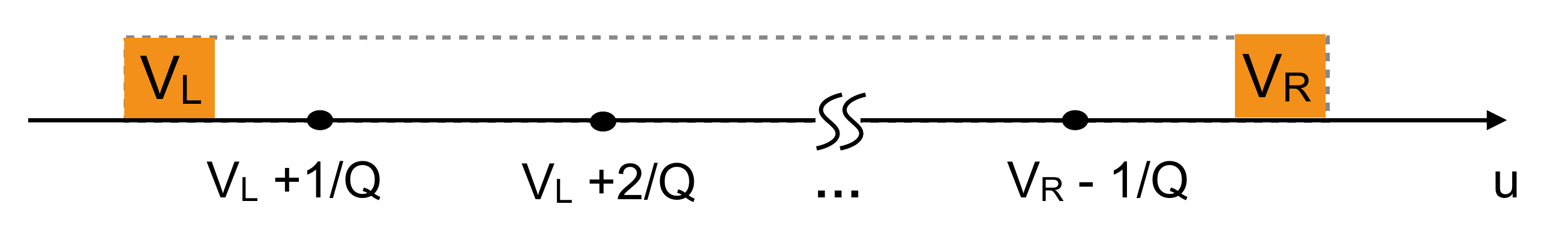}
   \caption{MV navigation segment $m:\left\{ l_L, l_R  \right\} = \left\{ (v_L, r_L, s), (v_R, r_R, s)   \right\} 
$. All viewpoints in the range $]v_L,v_R[$ are reconstructed by representations $(l_L,l_R)$.}
  \label{fig:segmentMV}
   \end{center}
  \end{figure}

We introduce    a novel variable, namely the \emph{multi-view navigation  segment}  $m$ defined  as a pair of representations $l_R$ and $l_L$ as follows
$$
m:\left\{  l_L, l_R  \right\} = \left\{ (v_L, r_L, s), (v_R, r_R, s)   \right\} 
$$
with the constraints that all viewpoints $u\in [v_L, v_R[$ are reconstructed by the representations $(l_L, l_R)$, as illustrated in Fig.   \ref{fig:segmentMV}. We      evaluate the distortion of a MV navigation  segment for a client interested in the navigation window $w$ of the video content $c$ as follows  
\begin{align}\label{eq:MVdistortion}
D_{mc} =  \frac{1}{N_u} \sum_{\substack{u \in [v_L, v_R[ : \\ u \in w}} d_{u,c}(l_L, l_R)
\end{align}
where $N_u$ is the number of viewpoints in $[v_L, v_R[$.

\begin{figure}[t]
   \begin{center}  
 \includegraphics[width=.75\linewidth,draft=false]{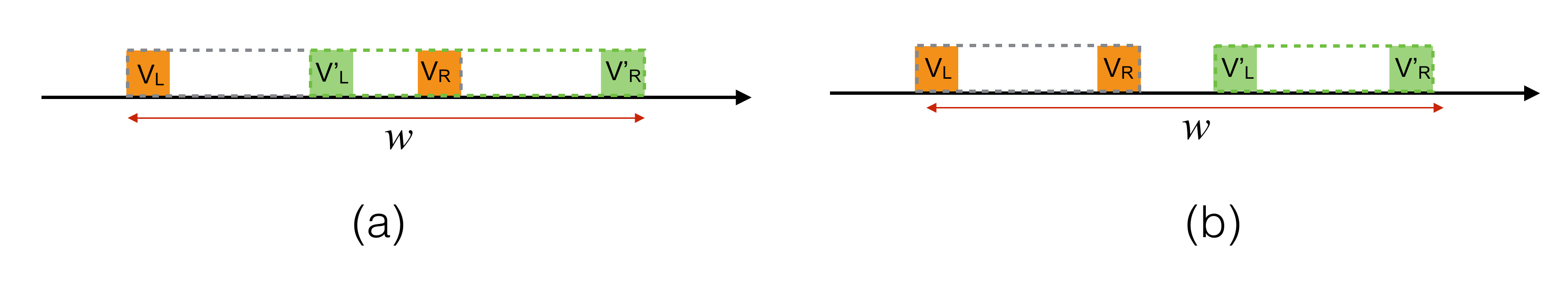}
   \caption{Example of two MV navigation segments $m:\left\{ l_L, l_R  \right\} = \left\{ (v_L, r_L, s), (v_R, r_R, s)   \right\}$ and $m^{\prime}:\left\{ l_L^{\prime}, l_R^{\prime}  \right\} = \left\{ (v^{\prime}_L, r^{\prime}_L, s), (v^{\prime}_R, r^{\prime}_R, s)   \right\} 
$ violating the constraint  {\bf C1} (a) and the constraint {\bf C2} (b). }
  \label{fig:constraints_violeted}
   \end{center}
  \end{figure}

We  consider  that  the client downloads a set of  MV navigation segments $\mathcal{M}$ instead of free set of representations that cover the navigation window. We denote by $R_{\mathcal{M}}$ the cost of downloading $\mathcal{M}$, which is the sum of the size of each representation in each MV navigation segment in $\mathcal{M}$.
For example,   a client that downloads  three representations $l_1, l_2,$ and $l_3$   associated to the camera views $v_1, v_2,$ and $v_3$, respectively (with $v_1 < v_2 <v_3$), has its data represented  by the segments $m:\left\{  l_1, l_2\right\}$ and $m^{\prime}:\left\{  l_2, l_3 \right\}$, and therefore   $\mathcal{M}=\{m, m^{\prime}\}$ and $R_{\mathcal{M}}=r_{l_1}+r_{l_2}+r_{l_3}$. 
A set of MV navigation segments $\mathcal{M}$ can be downloaded by a client if it meets the following constraints.
\begin{description}
\item [{\bf C1: }]\emph{re-use of camera views constraint:}   the  right representation of  one MV navigation segment shares the same camera view  with  the following MV navigation segment, with the exception of lateral segments. This means that neighbouring MV navigation segments share one  anchor view but possibly with different  coding rates.   
\item  \item [{\bf C2: }] \emph{full coverage constraint:} given a set of   MV segments downloaded by the client, all viewpoints in the client navigation window  $w$ need  to be covered by at least one MV navigation segment.  More formally, $\forall u\in w $,  $\exists m:\left\{  l_L, l_R  \right\}=\{(v_L, r_L, s), (v_R, r_R, s)  \} \in  \mathcal{M} $ s.t. $v_L\leq u < v_R$. 
 \end{description}
   In summary, the constraint  {\bf C1} imposes no overlap in the viewpoint domain among the MV navigation segments downloaded by the user. While the constraint  {\bf C2} forces the downloaded MV navigation segments to be contiguous. In Fig.   \ref{fig:constraints_violeted}, we depict two MV navigation segments violating the constraints   {\bf C1}  and   {\bf C2}. 

\subsection{MV  Navigation Segment Based Optimization}
We now  propose an  effective  algorithm to solve  the problem in \eqref{eq:problem_form} adopting the   new concept of MV navigation segment.  Given a set of representations  $\mathcal{T}$ available at the server, a client displaying a video content $c$ with a navigation window $w$ downloads the best set of MV navigation segments $\mathcal{M}$ defined as the one minimizing the distortion over $w$.  More formally, the problem in \eqref{eq:mean_dist} becomes
\begin{align}\label{eq:MVdistortion2}
 {D}^{\text{MV}}_{c}(w,  \mathcal{T}, B) = \min_{\substack{ \mathcal{M}:  l_L, l_R\subset \mathcal{T}  \\ R_{\mathcal{M}} \leq B  }} \sum_{m\in\mathcal{M}} D_{mc}(w)  
\end{align}
and the  optimization problem in \eqref{eq:problem_form} therefore becomes 
\begin{align}\label{eq:problem_form_2}
\mathcal{T}^* : &
\arg \min_{\mathcal{T}} \sum_{i\in \mathcal{I}} \sum_{u \in  \mathcal{U}_{c_i}}    \zeta_{i} q_{i}(u)
 \left\{   {D}^{\text{MV}}_{c_i}(w(u),  \mathcal{T}, B_i)\right\}
\nonumber \\
&\text{s.t. }      \sum_{l\in \mathcal{T}}  r_{l} \leq  |\mathcal{C}| C  
\end{align}
  The introduction of the MV navigation segments under the constraints {\bf C1} and  {\bf C2} limits the selection of possible anchor views   for a given range of virtual views. This simplifies the minimization problem in \eqref{eq:MVdistortion2}, which is considered in the overall optimization problem in \eqref{eq:problem_form_2}. As a result, in the special case of only one type of users with infinite bandwidth the representation selection problem can be solved in polynomial time \cite{toni2015network}.      In  a more general case, the   problem can still solved in a tractable way, as shown in the following subsection.  
   Finally, we comment on the equivalence   between \eqref{eq:problem_form} and  \eqref{eq:problem_form_2} under the constraints   {\bf C1} and  {\bf C2}. The latter constraint rules out cases in which one or more viewpoint $u$ cannot be reconstructed by any anchor view in the downloaded MV navigation segments.  No optimal set of representations   leaves a viewpoint in the navigation window with no anchor views ,  in the original problem in \eqref{eq:problem_form}, as it would lead to a zero satisfaction for the considered viewpoint. Therefore, both  the problems  \eqref{eq:problem_form} and  \eqref{eq:problem_form_2} have   solutions that cover the full set of viewpoints. On the contrary, the constraint  {\bf C1} excludes cases which could still lead to an optimal subset $\mathcal{T}^*$. However, it can be shown that in most of MV scenes, imposing the constraint   {\bf C1}  does not prevent to reach   optimality \cite{toni2015network}.  
 In Appendix \ref{sec:proofOpt}, we provide further details on the optimality conditions of the proposed algorithms.

\begin{ilp*}[h!]
\caption{\textbf{ }}
\begin{subequations} \label{eq:problem_form_ILP_2}
\begin{align}
  \min_{ \pmb{\alpha}, \pmb{\beta}, \pmb{\gamma}}& \sum_{i\in \mathcal{I}} \sum_{u \in  \mathcal{U}_{c_i}}    \sum_{c \in \mathcal{C}} 
\sum_{s\in\mathcal{S}}   \sum_{m:l_r,l_L\in \mathcal{L}}   \zeta_{i} q_{i}(u) \alpha_{icsm} \, f_{im}
  \label{obj_function} \\   \label{constr_b}
 \text{s.t. }&
 \alpha_{icsm} b_{csvm} \leq \gamma_{icsv}   &\forall i\in \mathcal{I},\forall  c\in \mathcal{C},\forall  s\in \mathcal{S}, \forall m\in \mathcal{M}, \forall v\in \mathcal{V}_c 
\\ \label{constr_c} &
 \gamma_{icsv}  \leq  \sum_i \sum_l b_{csvm} \alpha_{icsm}     &\forall i\in \mathcal{I},\forall  c\in \mathcal{C},\forall  s\in \mathcal{S}, \forall m\in \mathcal{M}, \forall v\in \mathcal{V}_c
\\ \label{constr_d} &
  \gamma_{icsv}   \leq   \beta_{csv}  &\forall i\in \mathcal{I},\forall  c\in \mathcal{C},\forall  s\in \mathcal{S}, \forall v\in \mathcal{V}_c
\\ \label{constr_e} &
 \beta_{csv}  \leq  \sum_i \sum_l   \gamma_{icsv}   &\forall i\in \mathcal{I},\forall  c\in \mathcal{C},\forall  s\in \mathcal{S}, \forall v\in \mathcal{V}_c
\\ \label{constr_f} &
\sum_m  a_{csum} \alpha_{icsm}  =\begin{cases} 1 &\mbox{if } c=c_i, s=s_i \\ 
 0  & \mbox{otherwise }  \end{cases}    &\forall u\in w_i, \forall i\in \mathcal{I},\forall  c\in \mathcal{C},\forall  s\in \mathcal{S}
\\ \label{constr_g} &
\sum_v    \gamma_{icsv}  r_{csv}  \leq B_i    &\forall i\in \mathcal{I},\forall  c\in \mathcal{C},\forall  s\in \mathcal{S}
\\ \label{constr_h} &
\sum_v \beta_{csv}  r_{csv}  \leq |\mathcal{C}| C    &\forall  c\in \mathcal{C},\forall  s\in \mathcal{S}
\\  \label{constr_i}&
  \alpha_{muic}  \in[0,1]  &\forall  m\in \mathcal{M}, \forall u\in \mathcal{U}_c, \forall i\in \mathcal{I}, \forall c\in \mathcal{C}
\\ \label{constr_j} &
      \beta_{csv}   \in[0,1]  &\forall  c\in \mathcal{C},\forall  s\in \mathcal{S}, \forall v\in \mathcal{V}_c
\\ \label{constr_k}&
    \gamma_{icsv}   \in[0,1]  &\forall i\in \mathcal{I}, \forall  c\in \mathcal{C},\forall  s\in \mathcal{S}, \forall v\in \mathcal{V}_c
\end{align}
\end{subequations}
\end{ilp*}
\subsection{ILP Optimization Algorithm} 
We now provide the ILP formulation that permits to solve the optimization  problem in  \eqref{eq:problem_form_2}. 
We introduce the following binary decision variables:
\begin{itemize}
\item
$ \alpha_{icsm}  = \begin{cases} 1 &\mbox{if users of type } i \mbox{ select  the MV segment } m \\
   						&  \mbox{ at resolution $s$ for video c    }  \\  
0 & \mbox{otherwise}  \end{cases} 
$
\item
$
\beta_{csv} = \begin{cases} 1 &\mbox{if camera view }  v \mbox{ for video c and resolution $s$ } \\
   						& \mbox{is  selected at least by one type of user } \\ 
0 & \mbox{otherwise}  \end{cases} 
$
\item
$\gamma_{icsv} = \begin{cases} 1 &\mbox{if camera view }  v \mbox{ for video $c$ and resolution $s$ } \\
  							& \mbox{is  selected  by  users of type  }  i \\ 
0 & \mbox{otherwise}  \end{cases} $
\end{itemize} 
We also define the following auxiliary matrices:
\begin{itemize}
\item
$a_{csum} = \begin{cases} 1 &\mbox{if viewpoint  } u \mbox{ is covered by } m \\ 
0 & \mbox{otherwise}  \end{cases} 
$
\item
$
b_{csvm} = \begin{cases} 1 &\mbox{if camera view }  v \mbox{ is a reference view in  } m \\ 
0 & \mbox{otherwise}  \end{cases} 
$
\end{itemize}
Finally, we also consider the matrix with entries $f_{im}$ defined as the     satisfaction of users of type  $i$  over   the segment of $m$, as described in Section \ref{sec:quality}.

With these variables, the optimization problem can be formulated as shown in \eqref{eq:problem_form_ILP_2}.   The objective function  \eqref{obj_function}  maximizes the sum of the user satisfactions while navigating over the scene of interest.  The constraints  \eqref{constr_b}-\eqref{constr_e}  set up a consistent relation between the decision variables $ \pmb{\alpha}, \pmb{\beta}, $ and $\pmb{\gamma}$. The constraint  \eqref{constr_f} imposes both the   constraints  {\bf C1}   and   {\bf C2}. Then,  the conditions \eqref{constr_g} and  \eqref{constr_h} reflect the client's bandwith constraint and the overall storage constraint, respectively. The constraints  \eqref{constr_i},   \eqref{constr_j} and   \eqref{constr_k} finally  force the decision variables to be binary ones.

The number of variables are of the order of $\mathcal{O}(|\mathcal{R}|^2V^2 |\mathcal{I}|  |\mathcal{C}|  |\mathcal{S}|)$, in the case  where all videos have the same number of camera views, i.e.,  $V_c=V, \forall c$. It is worth noting that   we defined classes of users as well as classes of video contents in our problem, bounding $|\mathcal{I}| $ and $ |\mathcal{C}|$  to small values.      Moreover, with a few  constraints that look reasonable in practice, such as  imposing   $|r_L-r_R|=n \Delta$, i.e., limiting the coding rate mistmatch between left and right anchor view \cite{de2015optimal},    we can narrow down even further  the number of variables to     $\mathcal{O}(V^2|\mathcal{R}| n  |\mathcal{I}|  |\mathcal{C}|  |\mathcal{S}|)$. Finally, since sparse camera arrangements are usually considered in practice, the value of $V$  is also bounded. Therefore, the proposed ILP problem in \eqref{eq:problem_form_ILP_2}  is solved with a tractable complexity.  
   
%
%

\section{Simulations}
\label{sec:settings} 
We now evaluate the performance of the proposed multi-view representations selection optimization. We have used the generic solver IBM ILOG CPLEX \cite{CPLEX} to solve different instances of the proposed ILP. We have considered different system configurations in our study that are  not meant to exhaustively cover all possible cases, but rather to illustrate the optimal coding strategy in several realistic cases. We first   describe the specific settings that we use to simulate the  illustrative and yet representative scenarios. Then, we compare the optimal     multi-view representation set with those computed with  baseline algorithms and illustrate the benefit of our solution.

 

\subsection{Simulation Settings}
\subsubsection{Satisfaction Function}
\label{sec:quality}
We first define the satisfaction function $f_{im}$ used in our ILP formulation as the satisfaction of a user of type $i$ when  he downloads  the segment $m$ of the video content of interest $c_i$. This is defined as     $f_{im}= 1- D_{mc_i}$  given by \eqref{eq:MVdistortion} with 
\begin{align}
\label{eq:distortion_DIBR}
 d_{u,c}(l_L, l_R) &=     \alpha D_{min} + (1-\alpha) \beta D_{max}  + \left[ 1-  \alpha  -  (1-\alpha) \beta \right] D_{I}
\end{align}
where $D_{min}=\min\{D_{l_L} ,D_{l_R}\}$, $D_{max}=\max\{D_{l_L} ,D_{l_R}\}$, $D_I$ is the inpainted distortion, and   $D_{l_L}$ and $D_{l_R}$ are the distortion of the left and right anchor view in the segment $m$.  We also define  $\alpha = \exp\left(-\xi |u-v_{min}| \right), $ and $\beta = \exp\left(-\xi |u-v_{max}| \right)$ 
with   $v_{min}=v_{l_l}$ if $D_{l_L} \leq D_{l_R}$, $v_{min}=v_{l_R}$ otherwise, and $v_{max}=v_{l_l}$ if $D_{l_L} >D_{l_R}$, $v_{max}=v_{l_R}$ otherwise.  The model is inspired by  \cite{toni2015network} 
 and the parameters $\xi$ and $D_I$, which  depend on the scene geometry, are computed by curve fitting.

Finally, we need to define  $D_{l}$ that is the distortion (due to coding artifacts) of the representation $l:(v_l, r_l,s_l)$ that is encoded at rate $r_l$.   We   model  $D_{l}$ as a \ac{VQM} score~\cite{VQM_software}, which is a full-reference  metric that has higher correlation with human perception than other MSE-based  metrics, as  shown in~\cite{Bess:C13}.   We can  evaluate this distortion as  
\begin{align}\label{eq:RD_model} 
D_l = D(r_l)  =     a -  \frac{b}{ r_l + e }
\end{align}

\begin{table}[t]
\begin{center}
\begin{tabular}{|c||c|c|c|c|c|}
  \hline
Video  		 &  $a$  	  & $b$   &   $e$   \\
  \hline   
 Shark 		&     $1$		&  $46.67$  	&  $-95.40$   \\  
 Undo Dancer 	&     $0.98$ 	&  $364.45$  	&  $868.08$   \\
 Hall	 		&    $0.98$ 	 & $83.57$  	&  $67.35$   \\
  \hline
  \end{tabular}
\end{center}
\caption{Fitting parameters for \eqref{eq:RD_model}.}\label{tab:fitting_param}
\end{table}
where $a$, $b$, and $e$ are parameters that depend both on the content characteristics     and the resolution   of the video. To evaluate the  parameters in \eqref{eq:distortion_DIBR}  and   \eqref{eq:RD_model}   for   representative categories of potential video sequences  in adaptive streaming systems, we considered three  multi-view video sequences at  $1080p$ resolution, namely ``Hall" (``movie" type of video), ``Shark" (``cartoon" type of video), and ``Undo Dancer" (``sport-action" type of video). The sequences  are highly heterogenous in terms of coding and view synthesis efficiency, and therefore they are representative of various video categories.  In Appendix \ref{sec:videocontent}, we provide further details on the characteristics of the video categories.

Note that    the above satisfaction model    is simple and yet accurate in   capturing the essential behaviors of the coding and synthesis artifacts. 
 More content- or scene-dependent models (e.g.,  the one in \cite{Hamza:C16})  that are usually more precise at the price of more   parameters to set up, can however be used as alternative in  our problem formulation, which is generic with respect to the quality metric.

 \begin{table} 
\begin{center}
\begin{tabular}{|c||c|c|c|c|c|}
  \hline
Connection  		 &  Wifi   & ADSL-fast & FTTH  \\
  \hline   
$B_{\text{min}}$ ($Mbps$) & 	 $0.4$  		&    $0.7$ 		&   $1.5$		\\
$B_{\text{max}}$ ($Mbps$) & 	 $4$ 	&   $10$ 		&   $25$		\\
connection index ($n$) & 1 & 2 &3 \\
Probability ($p_n$) & 	 0.4  & 0.3 & 0.3\\
  \hline
  \end{tabular}
\end{center}
\caption{Connection types.}\label{tab:connection}
\end{table}

\subsubsection{Clients Population}
\label{sec:clients}
We now form   users population   in order to have  a   representative set of potential clients in adaptive streaming services. In particular, we define how users types are characterized, and how the distribution of these types   of users is evaluated.

We consider    three    {\bf types of video content $c$} that users can select  with a uniform probability of  $1/3$,     as in the case of  most popular type of videos.     All videos have a camera set $\mathcal{V}=\{0, 8,16 \ldots, 72\}$ and a spatial resolution  of $1080p:(1920\times 1080)$, which is also the    display resolution of the users.
   Furthermore, we make the plausible assumption  that  the video content drives the focus of attention of the users  from which a navigation pattern can be evaluated. We therefore  define the probability for a user displaying content $c$ to request a navigation window centered on view $u$, i.e.,  $Q_{c}(u)$, as a Normal random variable with mean value $\mu_c$ being the mean value of the focus of attention and variance $\sigma_c^2$, i.e.,     $Q_{c}(u)\sim \mathcal{N}(\mu_c, \sigma_c^2)$. Without loss of generality, we assume the mean focus of attention to be the centre  of the camera views for all videos, i.e., $\mu_c = (v_V-v_1)/2$, while we assign different variance values to different video sequences.\footnote{ In our model, what is important is to understand if users are focused on similar navigation windows (homogenous scenario) or very different navigation windows (heterogenous scenario). It is therefore less important where the mean focus of attention is. It is rather much more important the variance from that mean value. For this reason, we assume that all sequences have the same $\mu_c$.} Namely, $\sigma_c^2 =\{80, 250, 3000\}$ for  ``Undo Dancer", ``Shark",  and ``Hall", respectively. For  the ``Undo Dancer" sequence, which represents a sport movie, most of the users will follow the main subject of interest (the dancer) and   most of the navigation windows  will overlap.  A small value of $\sigma_c^2$ is therefore considered. On the contrary, a larger value of $\sigma_c^2$ characterizes sequences where there are many focuses of attention and the navigation windows of users can   barely overlap. Finally, for a user of type $i$ displaying video content $c_i$,  $q_i(u)$,  the probability  of requesting a {\bf navigation windows} centered in $u$  is given by $q_i(u)=Q_{c_i}(u)$.   In our simulations, for each type of users we generate four possible navigation windows, extracted as realizations of the random process described above.

We then   characterize the    {\bf connection type}   that can be experienced by the users.  We define three types of connections, as provided in Table \ref{tab:connection} that is derived from   \cite{toni2015optimal}. The minimum and maximum bandwidth for downloading the representations of one chunk  are set  $2 B_{\text{min}}$ and $2 B_{\text{max}}$, respectively.  The probability $ p_n$ for a user to experience a given connection type $n$ is further provided  in the last raw of Table \ref{tab:connection}. For each video content, and for each connection type, we consider two types of users: $1)$ users  with $B_i$ set to the $25$th percentile of the available bandwidth for the considered connection, and $2$)  users  with $B_i$ set to   the $75$th percentile of the available bandwidth for the considered connection. We assume a probability $1/2$ for a user to  experience one or the other downloading bandwidth. As a result, we have with our representation $18$ types of users that are representative of realistic scenarios.  
The video content $c_i$ and the experienced bandwidth $B_i$ identify  the users of type $i$ with  probability $\zeta_{i}= (1/|\mathcal{C}|)  (1/2)p_{n_i}  = p_{n_i}/6$, with ${n_i}$ being the connection  of the type of user $i$.

\subsubsection{   Comparative  Algorithms}
We compare the representation sets selected by our algorithm  to  the recommended sets of Apple, Netflix, and YouTube~\cite{Apple,Netflix_blog,Kreuzberger:C16}. 
In particular,  we consider the bitrate recommended for the $1080p$ resolution, which means  $\{11, 24, 39\}  \, Mbps$ for Apple,  $\{4.3,5.8\}  \, Mbps$ for Netflix,  and  $\{4.072\} \,  Mbps$ for YouTube. To guarantee that at least one pair of anchor views can be downloaded also for the users experiencing poor channel connections,  we   add a representation encoded at $400  \, kbps$ to all three data sets. From these available encoding rates for all available camera views, we optimize the subset of views that best satisfy the clients population.

 To better evaluate the benefit of a \emph{complete adaptation}  of the representation set to content characteristics, users' connections  as well users' interactivity levels, we introduce a comparative algorithm that adapts to only  part of this information. More in details,  we introduce this comparative algorithm as a possible extension of the one proposed  in \cite{toni2015optimal}  to MV setting. Basically,  the  representation  set  is optimized in such a way that it adapts to the   population and content information, but   not necessarily to the interactivity aspect. This methods  first selects a subset of camera views to be stored   at the server and then   selects the encoding bitrates for    selected camera views, imposing the same coding rate for all camera views.  For the camera view selection,  we   assume a regular sampling of the  views, with  camera sampling values  $L= 8, $ and $16$. The selection of the bitrates is then optimized with  our ILP optimization problem, but with the additional  constraint of imposing the same coding bitrates to the views.   
 We label this optimization as ``PA" (\emph{partial adaptation}) in the following.

\begin{figure} 
\centering
\subfigure[Camera spacing $8$]{ \includegraphics[width=.43\linewidth,draft=false]{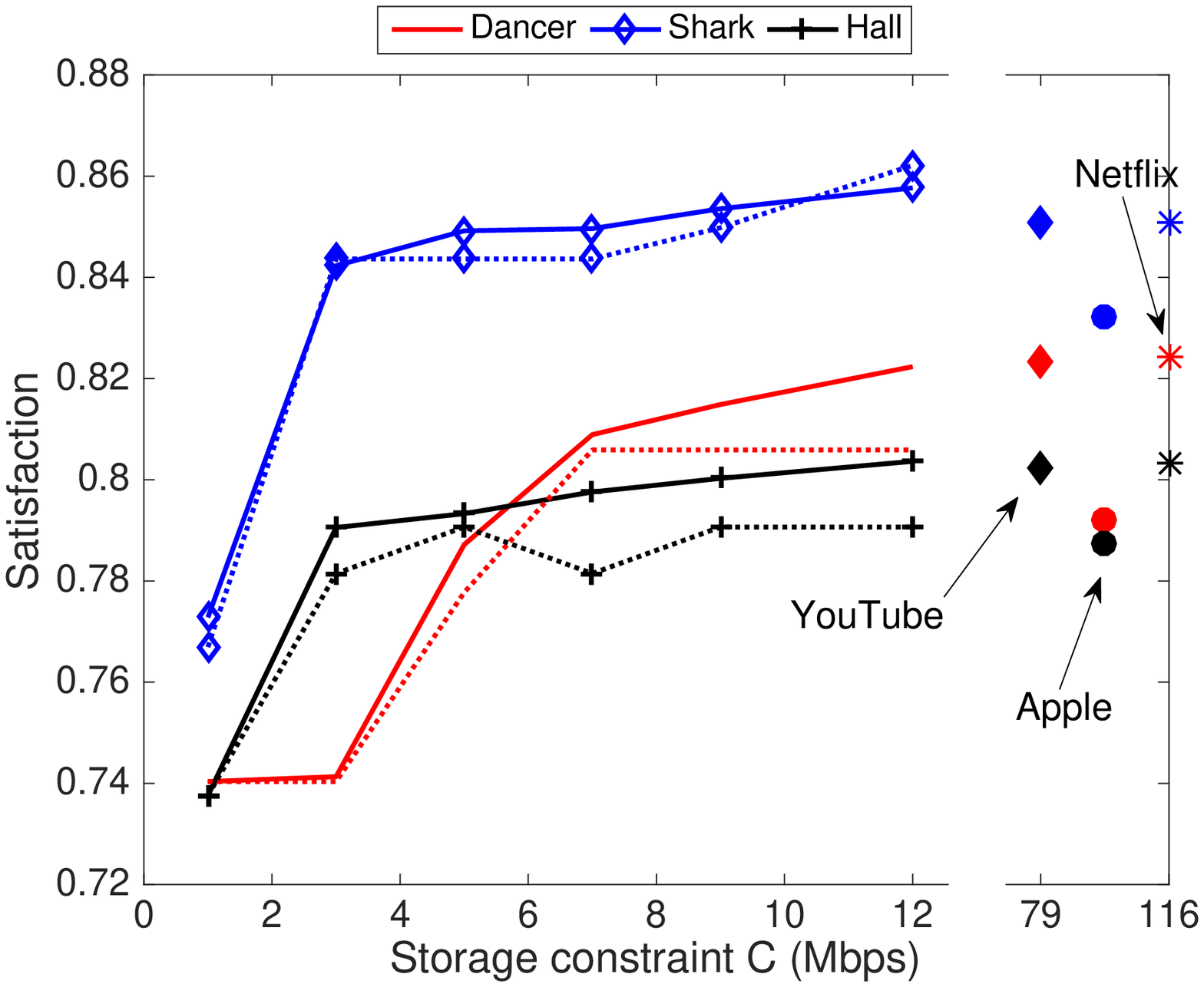}
\label{fig:Satisfaction_perVideo_Nv_3_Spacing8}}
\hfill
\subfigure[Camera spacing $16$]{ \includegraphics[width=.47\linewidth,draft=false]{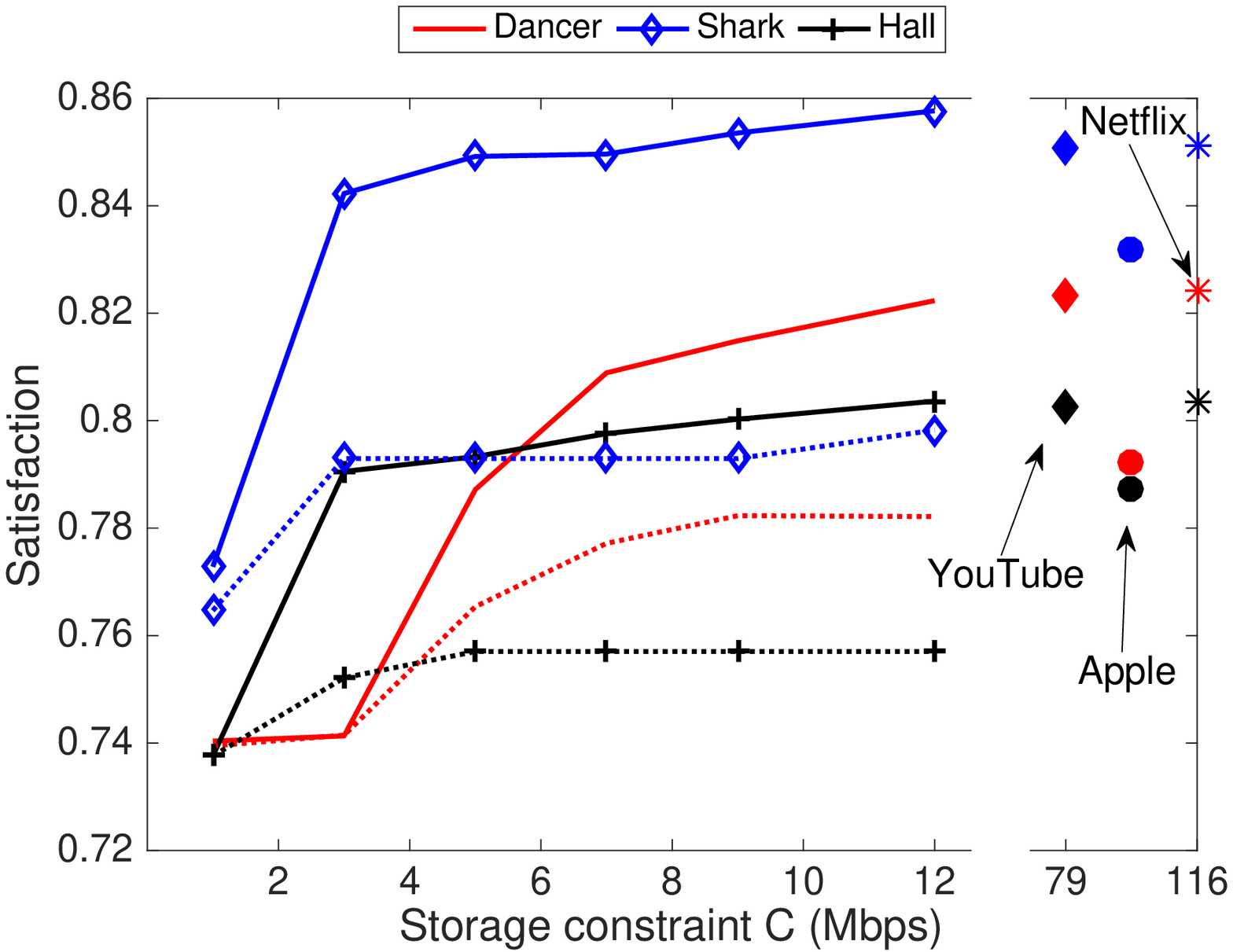}
\label{fig:Satisfaction_perVideo_Nv_3_Spacing16}}
\caption{Expected satisfaction per video vs.  the storage capacity constraint for the ``NW-homogenous" scenario.   . Solid lines show the performance of the optimal multi-view representation set proposed in this paper, while dashed lines show the performance of the PA  method. 
}
\label{fig:Satisfaction_perVideo__5bw_BASELINE}
\end{figure}

\begin{figure}
\centering
 \includegraphics[width=.55\linewidth,draft=false]{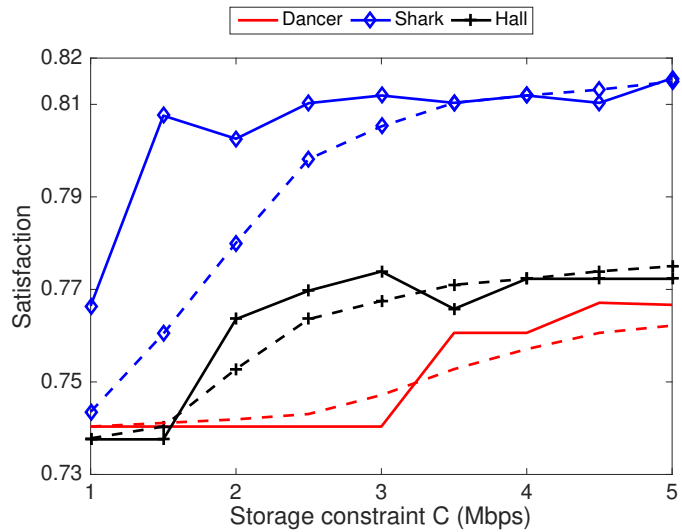}
\caption{Comparison between joint optimization of the representations of all videos  (solid line) and independent optimization per video (dashed line) for the ``NW-homogenous" scenario.      }
\label{fig:joint_vs_indp}
\end{figure}

\subsection{Simulation Results}
\label{sec:results} 
In the following, we first show how our optimal set of representations performs with respect to the competitor  algorithms, i.e., the PA algorithm   as well as the recommendations by Apple, YouTube, and Netflix. Then, we simulate realistic adaptive streaming clients and show the quality experienced over time for different clients both in the case of the optimal and commonly recommended set of representations.  

\subsubsection{Optimal set of representations}
In the first considered scenario (labeled ``NW-homogenous" scenario),  we assume that all video content have a navigation window   $w=[16,52]$ for all users, rather than a probabilistic model based on $Q_{c}(u)$. This means that  navigation windows are highly homogenous for all users. This setting penalizes less the competitor algorithms that do not optimize the encoding rates as well as the camera views with respect  to the interactivity of the users.    
In Fig.~\ref{fig:Satisfaction_perVideo__5bw_BASELINE}, we provide the expected satisfaction   of the clients  with respect to $C$, the mean storage capacity per video.   Simulation results are provided for   our optimal set of multi-view representations (solid lines) as well as for the representation set derived from  the PA algorithm (dashed lines) with  a  camera sampling values  $L= 8,16$  in Fig.~\ref{fig:Satisfaction_perVideo_Nv_3_Spacing8} and Fig.~\ref{fig:Satisfaction_perVideo_Nv_3_Spacing16}, respectively.   Note that $8$ is the minimum camera spacing considered also in our camera arrangement. Therefore,  in Fig.~\ref{fig:Satisfaction_perVideo_Nv_3_Spacing8},  the main difference between the proposed optimization algorithm  and the PO one is that the bitrate is selected a priori  for all camera views in the latter. In Fig.~\ref{fig:Satisfaction_perVideo_Nv_3_Spacing8}, we can notice the gain in the extra degree of freedom provided by our optimization. Note also that there is no a strictly  monotonic  increasing behavior of the satisfaction function   with the storage constraint $C$.      Denoting by $\mathcal{T}(C)$ and  $\mathcal{T}(C^{\prime})$ the two optimal set for constraints  $C$ and $C^{\prime}>C$, respectively, we do not necessarily have that  $\mathcal{T}(C) \subset \mathcal{T}(C^{\prime})$ in our integer optimization problem. The optimal set of representations can even  change substantially when the capacity constraint increases from $C$ to $C^{\prime}$. The satisfaction function might therefore have  slightly oscillating  behavior that corresponds to minor quality variations in practice.

When the camera sampling is  $L=16$, the performance  of the PA algorithm drops substantially with respect to our optimal set of representation, as shown in Fig.~\ref{fig:Satisfaction_perVideo_Nv_3_Spacing16}.    Note that in the literature, a VQM gain of $0.1$ is considered to be a significant improvement, and the gain that we achieve is about $0.05$ for the ``Shark" video sequence. Finally, we also provide the satisfaction level achieved by the representation set recommended by Apple, YouTube, and Netflix.  We notice that we achieve a better satisfaction level with a much lower storage capacity cost.  This not only leads to a gain in terms of storage cost but also in terms of CDN capacity. In \cite{toni2015optimal}, it has been shown that  a lower storage cost of the optimal set translates into a reduced CDN cost, which is usually a bottleneck in nowadays adaptive streaming systems.

  We now show  in more details the gain obtained  in  jointly optimizing the multi-view representations for several video types.   We optimize   jointly the  representations of all three contents   with a mean storage capacity of $3C$ and we compared it with the case in which   the representations of each video are optimized  independently from other video sequences with  a storage constraint of $C$ for each video. In both cases, the representation set is optimized with our proposed algorithm.  
   In Fig.~\ref{fig:joint_vs_indp}, we show the performance for both the joint and the independent optimization.  We can observe that when all representations are optimized jointly, there is a better usage of the storage capacity. This is due to the fact that an unequal allocation of the storage capacity among videos is beneficial in the case of different video characteristics. For example, videos with low complexity content can be encoded at low coding rates, in favor of more complex sequences that might occupy a storage capacity larger than the average capacity $C$.  Note also that the joint optimization does not always outperform the independent one for all videos. For example, the ``Dancer" sequence has a lower quality in  the joint optimization, but for a better quality of the other two video sequences, resulting in a better expected satisfaction   among all users.

 \begin{figure} 
\centering
\subfigure[$C=1 \, Mbps$]{ \includegraphics[width=.45\linewidth,draft=false]{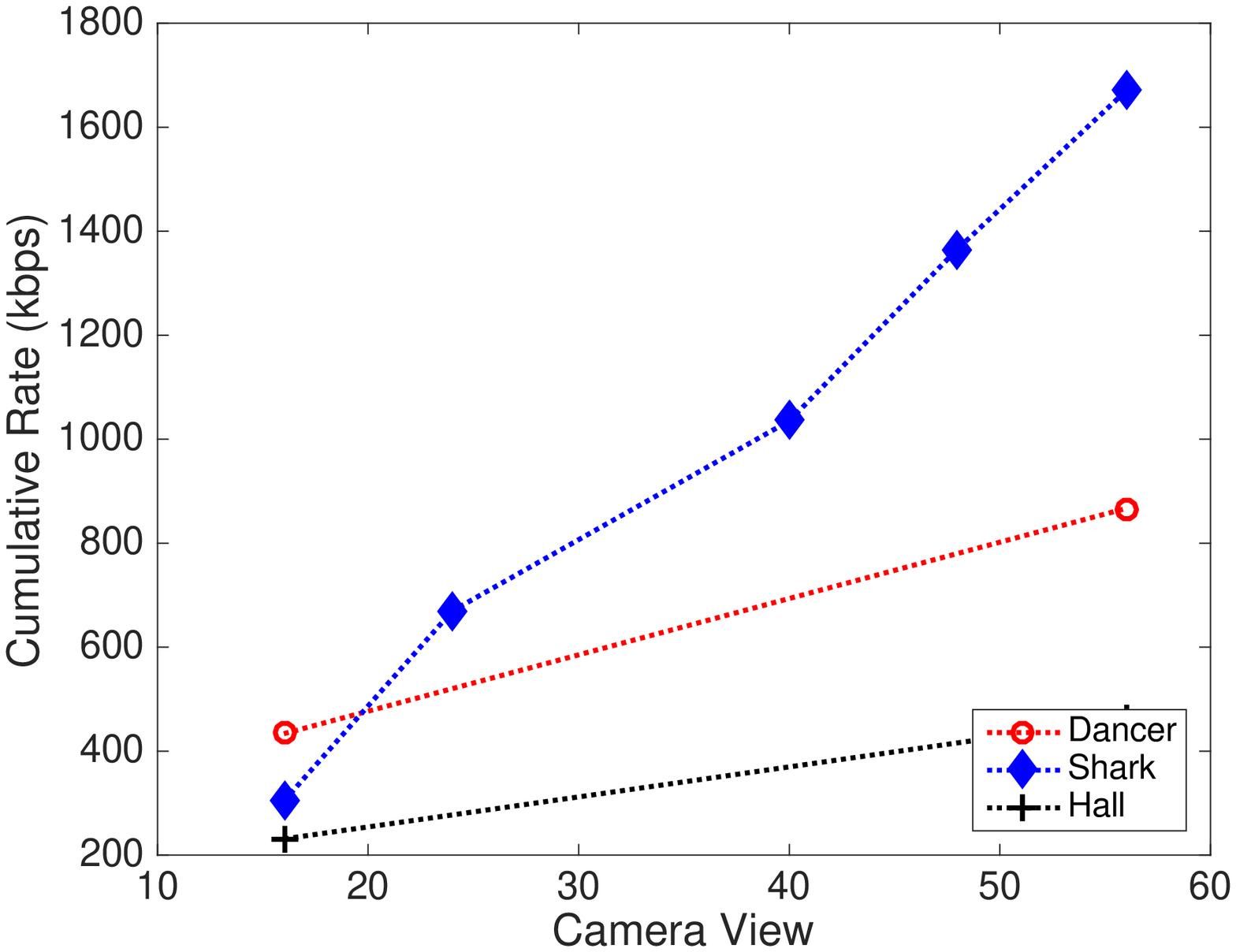}\label{fig:CUMsum_C1000_Nv_3_Scenario2}}
\hfill
\subfigure[$C=5  \, Mbps$]{ \includegraphics[width=.45\linewidth,draft=false]{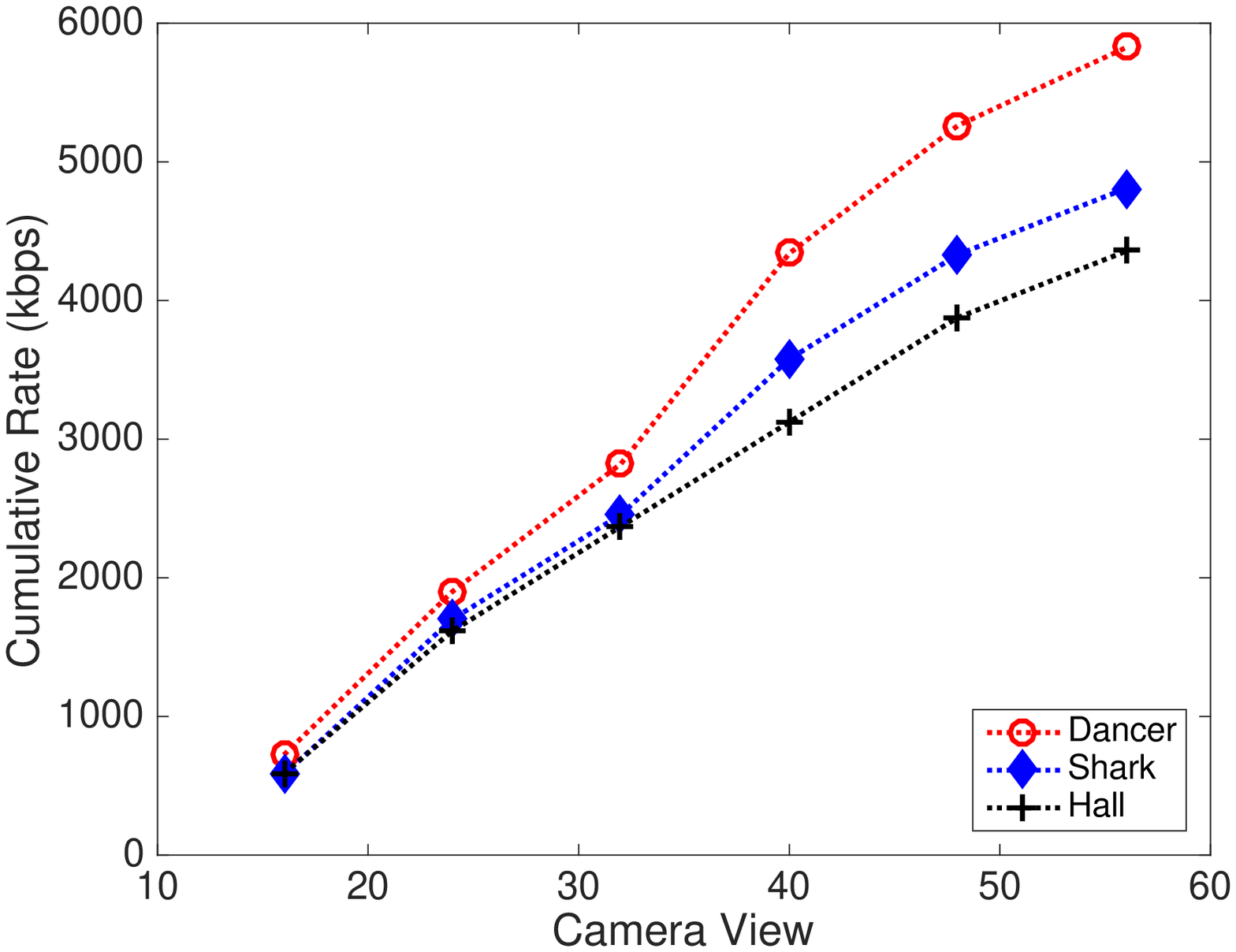}\label{fig:CUMsum_C5000_Nv_3_Scenario2}}
\hfill
\subfigure[$C=12  \, Mbps$]{ \includegraphics[width=.45\linewidth,draft=false]{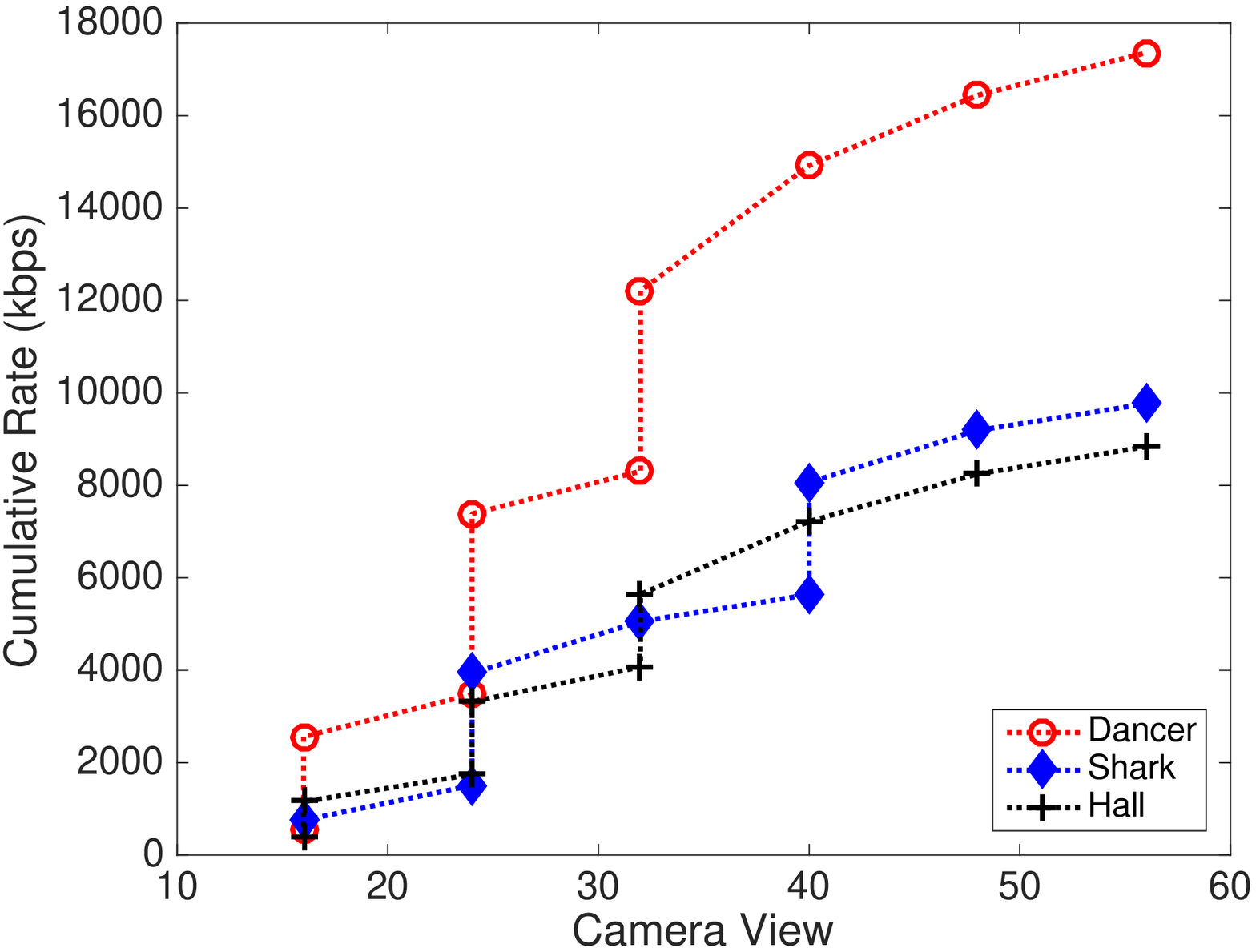}\label{fig:CUMsum_C12000_Nv_3_Scenario2}}
\caption{Selected optimal representation sets for different storage capacities for the ``NW-homogenous" scenario with a  navigation window $w=[16,52]$ for all users. }
\label{fig:representatinos_Nv_3_het}
\end{figure}

To better understand the unequal allocation of the storage capacity among different videos and different representations, we show how the    representations   achieved by our joint optimization for different video categories and storage constraints in   the homogenous scenario of  Fig.~\ref{fig:Satisfaction_perVideo__5bw_BASELINE}. 
In Fig.~\ref{fig:representatinos_Nv_3_het}, the cumulative sum of the optimal representation rates   is provided as a function of the camera view, for different videos.    Each point along the curves is an additional representation, whose view index is indicated in the x-axis and the cumulative rate (summing also the rate of the previous representations) is indicated in the   y-axis. 
In a highly constrained scenario with a storage constraint per video of  $C=1  \, Mbps$ (see Fig.~\ref{fig:CUMsum_C1000_Nv_3_Scenario2}),  only the lateral viewpoints $16$ and $56$ are stored for ``Dancer" and ``Hall" sequences, while multiple inner camera views are stored for the ``Shark" video sequence. This can be explained  by the fact that the latter is   affected by the synthesis process and slightly by the coding process, as shown in  Table \ref{tab:video_character}. Note that  $C=1  \, Mbps$ is a highly challenging scenario, therefore only one coding rate is provided per camera view, and in general it is a low coding rate\footnote{The adaptive streaming characteristic is preserved in realistic DASH systems since $i)$ extra storage capacity is dedicated to  different spatial resolution sizes; $ii)$ the value of $C$ can be increased till most of the views have several representations.}.
 Increasing the storage constraint to $C=5  \, Mbps$  (see Fig.~\ref{fig:CUMsum_C5000_Nv_3_Scenario2}),  more camera views are stored also for the other video sequences and the rate allocated per camera view  is similar among the three video sequences. This is however true only for this specific storage constraint.   
Increasing $C$ to $12  \, Mbps$ (Fig.~\ref{fig:CUMsum_C12000_Nv_3_Scenario2}), we notice that many camera views  are stored at the server at multiple coding rates  also for   the ``Dancer" sequence (camera views $16, 24,$ and $32$ for example are encoded at two different coding rates). This sequence  is the one  suffering  the  least from  view synthesis distortion, therefore it is has many views  only for large $C$ constraints. At the same time, it highly suffers from coding artifacts. For this reason,   the bitrates for the camera views of ``Dancer" are generally higher than the representations of the other videos.

\begin{figure} 
\centering
\subfigure[$C=1  \, Mbps$]{ \includegraphics[width=.4\linewidth,draft=false]{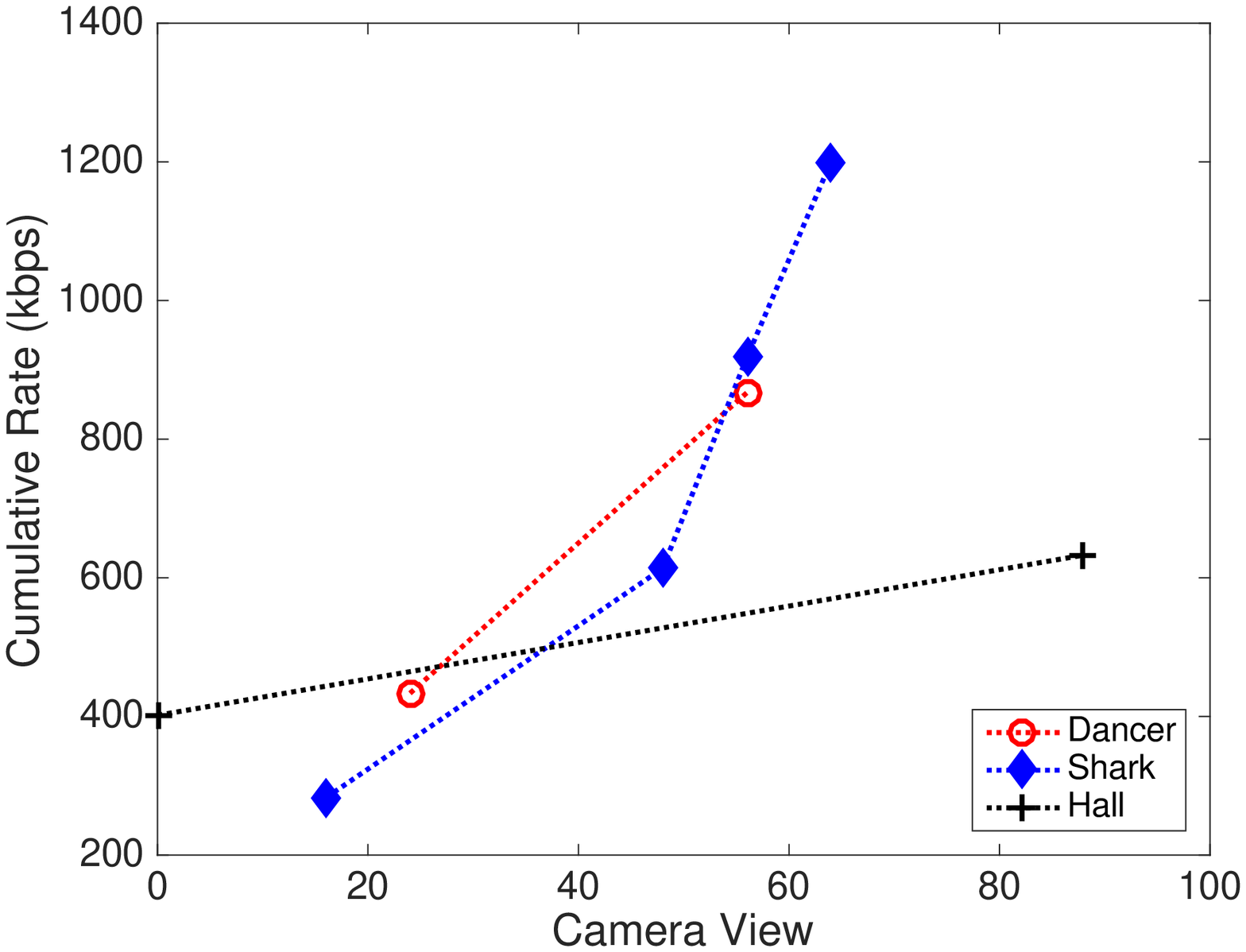}\label{fig:CUMsum_C1000_Nv_3_Scenario5}}
\hfill
\subfigure[$C=5 \, Mbps$]{ \includegraphics[width=.4\linewidth,draft=false]{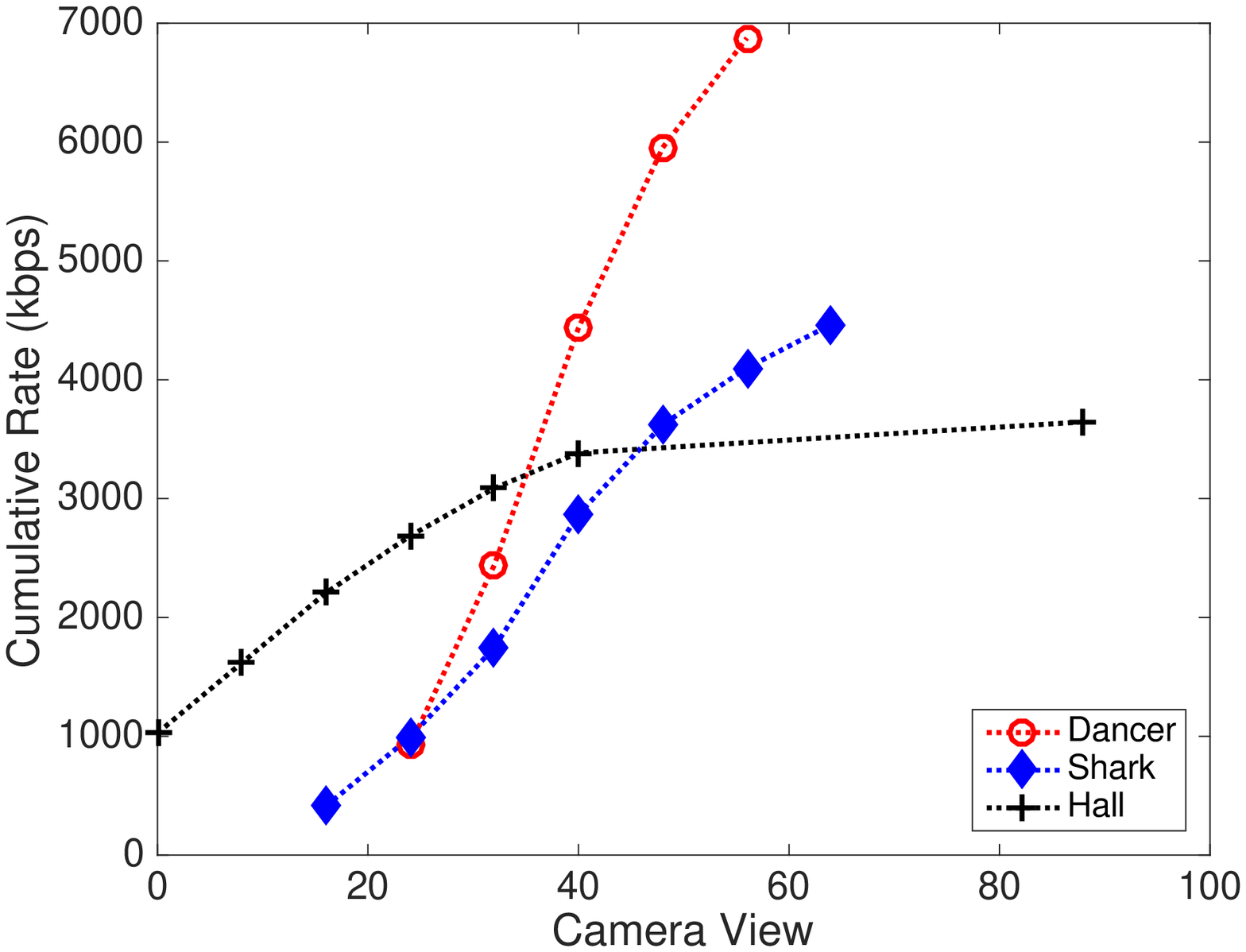}\label{fig:CUMsum_C5000_Nv_3_Scenario5}}
\hfill
\subfigure[$C=12  \, Mbps$]{ \includegraphics[width=.4\linewidth,draft=false]{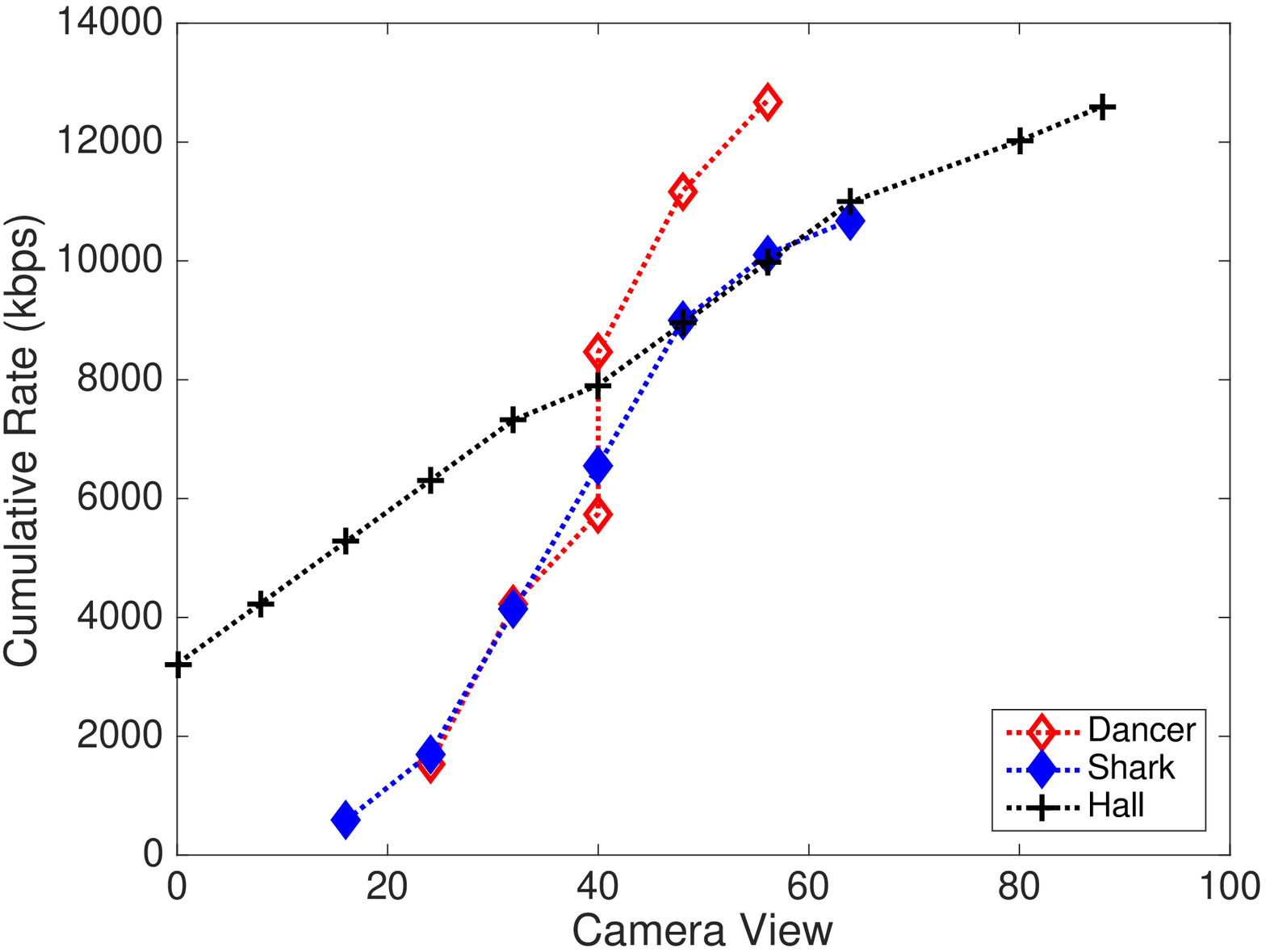}\label{fig:CUMsum_C12000_Nv_3_Scenario5}}
\caption{Selected optimal representation sets for different storage capacities for the ``BW-homogenous" scenario.   }
\label{fig:CUMsum_Nv_3_Scenario5}
\end{figure}

  We now consider a different scenario in which users are more heterogenous in terms of navigation paths but  more homogenous in terms of experienced bandwidth.  Namely, we set  $\sigma_c^2 =\{10, 150, 3000\}$ for  ``Undo Dancer", ``Shark",  and ``Hall", respectively, as previously described in Section \ref{sec:clients}, and $p_n=1$ for Wi-Fi connection and zero for the remaining ones.    This scenario, labeled ``BW-homogenous",  allows us to understand the effect of different interactivity levels on the optimal set of representations.   In Fig.~\ref{fig:CUMsum_Nv_3_Scenario5}, we provide the optimal representation set for the ``BW-homogenous". 
In the case of $C=1 \, Mbps$ (see Fig.~\ref{fig:CUMsum_C1000_Nv_3_Scenario5}), only  the sequence ``Shark" has more than two camera views encoded in the optimal set, similarly to what observed in the previous results. When the storage constraint $C=12  \, Mbps$  (se Fig.~\ref{fig:CUMsum_C12000_Nv_3_Scenario5}),  all sequences have more than two camera views encoded in the optimal set.   ``Undo Dancer"  has fewer  camera views ($6$ in total) but at high encoding rate, while   ``Hall" has more camera views (11 in total) since   a wider navigation window is required for this sequence. At the same time, since ``Hall" is not drastically affected by coding artifact, the camera views are encoded at medium rate. Finally, it is worth noting that very few camera are encoded at multiple rates. This is because in this specific scenario users are highly homogenous in terms of available bandwidth. 

These two very different simulated settings show the importance of optimally designing the representation set based on the video characteristics as well as the clients population. In particular, the above results  show that   unequal allocation of the storage capacity among video types as well as camera views is essential to strike for the right balance between storage cost and users satisfaction. This opens a large number of perspectives in free-viewpoint adaptive streaming, namely $i)$ it shows the importance to take into account the users' interactivity in the design of the representation set; $ii)$ it provides a theoretical framework that captures the complexity of today's video delivery systems\footnote{Note that by adding constraints the the ILP problem more aspects of the delivery systems (e.g., constraints during the delivery) can be included in our problem formulation. }.

%

\subsubsection{Experienced quality at the client side}
After we have illustrated    satisfaction functions that capture gains in terms of system design, we now provide results in terms of average quality experienced by realistic clients  navigating in a 3D scene.    We still optimize the optimal representation set with our proposed algorithm, but we evaluate the  performance differently, i.e., by simulating actual individual  users.      
In particular, we consider $U$ users, randomly assigned to one of the available classes. A user is associated to type $i$ with probability $\zeta_i$. Recall that the type  defines the video content of interest for the user, but also the type of connection in terms of      \emph{average} downloading rate. For each user,  we randomly generate the \emph{actual} downloading rate  over time.  The channel is generated assuming a Markovian channel as also considered in \cite{Zhou:J16,Chiariotti:2016}.  Finally, each user of type $i$ is associated to a starting viewpoint $u$ with probability $q_i(u)$. From this starting request, for each user, we randomly generate an  interactive session  over time, i.e., we simulate the user navigating in the scene over time following the model in  \cite{Toni:A14}.  Given the requested viewpoint $u$, each client evaluates the navigation window $w(u)$ that can be requested by the user during the $2s$ of the temporal video chunk.   The adaptation logic of the user defines the best set of cameras that cover the navigation window of interest.  This means that each user will request the  camera views among the representative set available at the server  that    minimize the distortion of the navigation window of interest and yet satisfy the experienced downloading rate. The optimization is performed following the dynamic programming  algorithm in \cite{toni2015network} and described in Appendix \ref{sec:adatLogic}.  This adaptation logic is inspired by \cite{Hamza:2014} where a two-step algorithm is proposed: it first selects the camera views to download    by predicting possible navigation paths of the user;   then the best representations for these views      are evaluated. In our case, we assume that   the navigation window is selected first   and then the best representations  (in terms of camera views and rate) are selected.

 \begin{figure} 
\centering
 \includegraphics[width=.55\linewidth,draft=false]{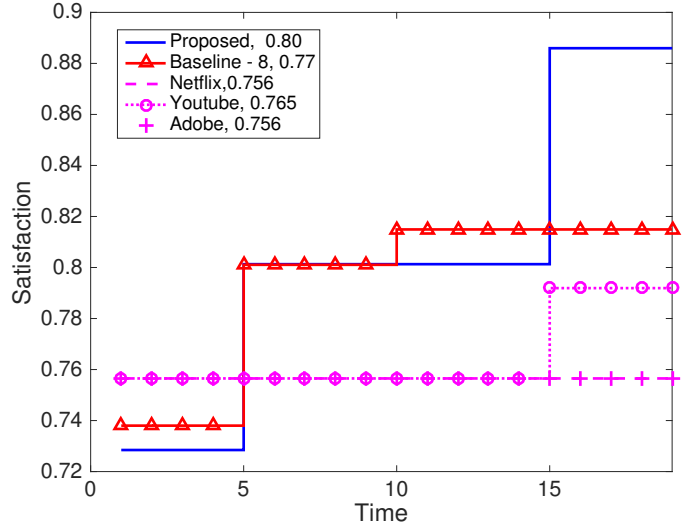} 
\caption{Mean satisfaction experienced by clients over time for a staircase bandwidth channel for the ``NW-homogenous" scenario.     The representation set of the proposed method is the one of   Fig.\ref{fig:CUMsum_C12000_Nv_3_Scenario2}.  Satisfaction values averaged over time are provided in the legend.}
\label{fig:Comparison_temporal_Scenario2C12k}
\end{figure}
 
We now  provide a first simulation for a scenario of $U=40$ users are simulated over $30$ time slots and averaged over $100$ loops to get statistically meaningful results.   
  Fig.   \ref{fig:Comparison_temporal_Scenario2C12k} depicts the average satisfaction per user when different representation  sets   at the server side have been optimized  for the case of $12  \, Mbps$  of storage capacity (per user).  For this figure, we assume a staircase channel that is constant for $5$ time slots and then increases. The experienced values over the $30$ time slots are $2  \, Mbps$,$4  \, Mbps$, and $6  \, Mbps$.
  We consider this specific channel behavior to better show the temporal evolution of the clients  satisfaction.    In the figure legend, we also provide the mean satisfaction over time.          It can be shown that the set optimized with the proposed ILP formulation achieves the highest satisfaction.  
  Looking at the PA   optimization method, we observe that, if the camera sampling is $L=8$, the algorithm performs yet quite well. However, compared to the recommended sets, the proposed solution achieves a substantial gain.

 \begin{figure} 
\centering
\subfigure[User downloading ``Undo Dancer" video]{ \includegraphics[width=.55\linewidth,draft=false]{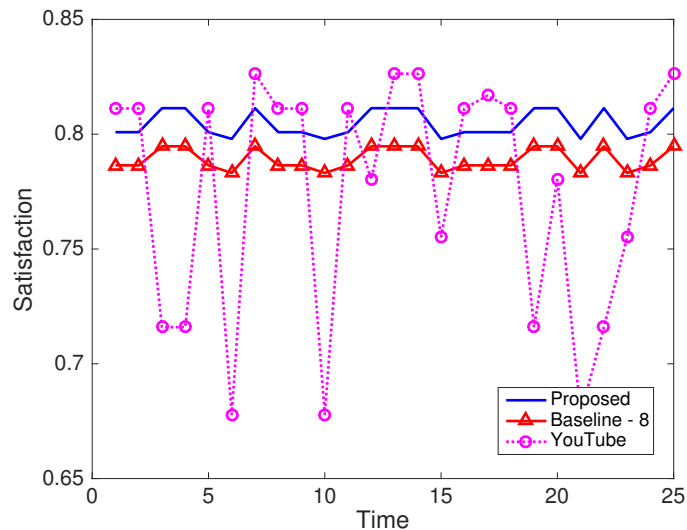}\label{fig:QvsTime_Sceanrio5_User2}}
\hfill
\subfigure[User downloading ``Hall" video]{ \includegraphics[width=.55\linewidth,draft=false]{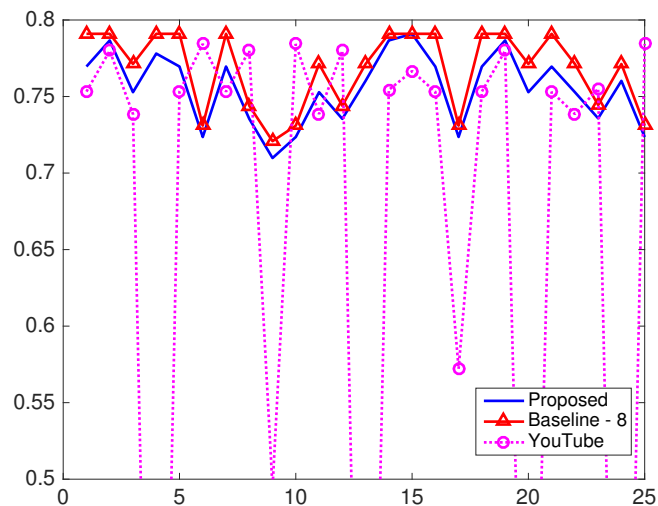}\label{fig:QvsTime_Sceanrio5_User26}}
\caption{Per user satisfaction over time for the ``BW-homogenous" scenario.}
\label{fig:QvsTime_Sceanrio5}
\end{figure}

In Fig. \ref{fig:QvsTime_Sceanrio5}, we   show the satisfaction function for the 
``BW-homogenous" scenario and we provide the per user satisfaction rather than the mean one. For the sake of clarity, we show the results only for the YouTube recommend set, since it is the best among the three competitor solutions under considerations. The results are provided for one representative user asking for ``Undo Dancer" and ``Hall" video sequences, respectively. For both cases, the   optimal set substantially outperforms the YouTube set. In particular, the YouTube data set can be sustained by the network only in few temporal instants. When it is not sustainable, a rapid loss of the satisfaction is experienced. When we compare     the proposed optimal set and the PA one, we observe that    the proposed optimization allocates a less-optimal representation set to the ``Hall" video sequence, to gain in terms of performance for the ``Undo Dancer" video. This is motivated by the fact reducing the number of representations stored at the server for   the ``Hall" video sequence reduces slightly the quality experienced for this content, but at the same time it allows that more representations are stored for the other sequences, substantially improving the quality of these contents.

This behavior can be better explained in the following plots. 
In Fig. \ref{fig:Qvsusers_Sceanrio2}, we provide the mean satisfaction per users in the case of $30$ users in total. The satisfaction is averaged over time and provided for both the proposed and the baseline optimal sets. We considered both the ``NW-homogenous" and ``BW-homogenous" scenarios. In both figures, users are ordered based on the requested videos. This means that the first $10$ asks for the ``Undo Dancer" video, the last $10$ asks for the ``Hall" video, while the inner users ask for ``Shark" video.  We also provide the mean satisfaction (average both over time and for all users)  in the figure legend.  In both scenarios, the proposed optimal set outperforms the baseline one in terms of overall mean satisfaction. However, looking at the quality at which each user displays the required video, we observe that for some users the quality perceived by the baseline set is better. For example, in the ``NW-homogenous" scenario, the ``Shark" video is better displayed when the baseline set is available. This gain however is paid by the other users, who suffer when downloading the  other two videos.

\begin{figure} 
\centering
\subfigure[``NW-homogenous" scenario.]{ \includegraphics[width=.55\linewidth,draft=false]{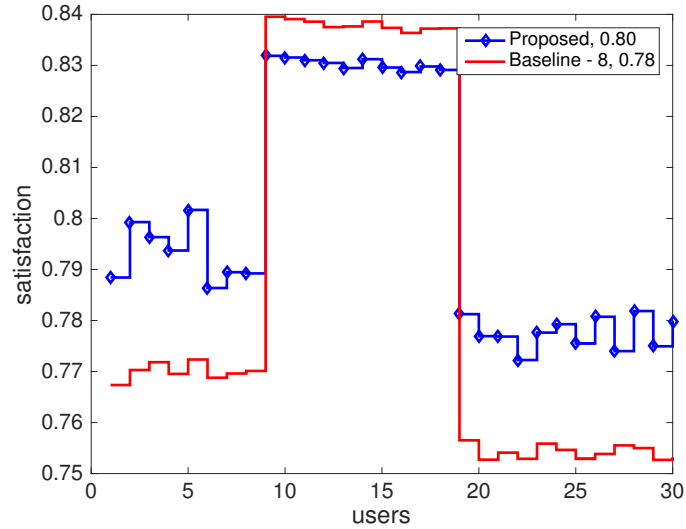}\label{fig:Qvsusers_Sceanrio2}}
\hfill
\subfigure[``BW-homogenous" scenario.]{ \includegraphics[width=.55\linewidth,draft=false]{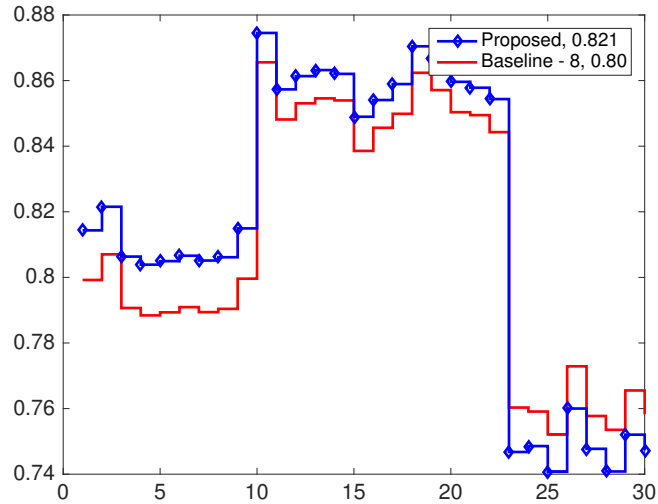}\label{fig:Qvsusers_Sceanrio5}}
\caption{Average (over time) satisfaction per user for a storage constraint $C=12$Mbps.}
\label{fig:Qvsusers_Sceanrio2}
\end{figure}

    In summary, we have shown that for simulated users in  interactive multiview video content  a better quality is achieved when the optimal representation set is designed following the proposed optimization. This leads to a better satisfaction achieved by the users, but also to a better usage of the storage space available at the server side. This better usage can also be reflected in  reduced CDN costs, which is a very important aspect  in nowadays adaptive streaming systems. 
    
\section{Conclusions}
To the best of our knowledge, this paper is the first study about  optimal encoding parameters for representation sets in free-viewpoint adaptive streaming. We have 
 defined an optimization problem for the selection of the representation set that maximizes the average satisfaction of interactive users while minimizing their view-switching delay. We  define a novel variable, namely the multi-view  navigation segment, and formulate an   optimization problem that can be   solved as a tractable   ILP   problem. We   characterize the satisfaction of interactive users as the quality experienced by the user during the navigation. This function is able to take into account both coding  and  view synthesis artifacts.     We  finally measure the performance of representation sets based on content provider recommendations  and show the suboptimality of baseline algorithms that do not adapt the coding parameters to the video and users characteristics. We therefore highlight  the gap between existing recommendations and solutions that maximize the average user satisfaction.    In particular, we show that an unequal allocation of the storage capacity among different video types as well as camera views is essential to strike for the right balance between storage cost and users satisfaction in interactive multi-view video systems.

 \bibliographystyle{IEEEtran}
\bibliography{All}

\appendices

\section{Video Content Characteristics}
\label{sec:videocontent}
\begin{figure} 
 \begin{center}
 \includegraphics[width=0.4\linewidth,draft=false]{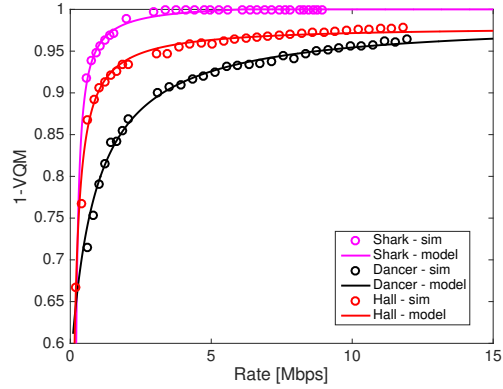}
 \caption{Satisfaction function vs encoding rate for different sequences encoded with HEVC codec. }
 \label{fig:VQM_Videos_1920p}
 \end{center}
 \end{figure}

We now better describe the video characteristic of the scene that we consider in our simulations settings. The sequences have been selected since they are highly heterogenous in terms of coding and view synthesis efficiency, and therefore they are representative of video categories.  All sequences have been encoded with HEVC \cite{sullivan2012overview} with an encoding range of $[600 \ kbps- 120\ Mbps]$ and  a resolution of $1080p$. 
 Different coded versions are then used as anchor  views for several virtual views. We then evaluate the quality of the reconstructed viewpoints as $(1-\text{VQM})$ score.    In Table \ref{tab:fitting_param}, the parameters $a$, $b$, and $e$ for    \eqref{eq:RD_model}   are provided for each of the considered sequences  and in  Fig.~\ref{fig:VQM_Videos_1920p}, we provide the quality as a   function of the encoding bitrate for the   considered video sequences,  for both the experimental VQM and the quality derived from \eqref{eq:RD_model} for a representative camera view.  By curve fitting, we then  evaluate the  parameters $\gamma$ and $D_I$  in  \eqref{eq:distortion_DIBR}.
We set the inpainting distortion to $D_I=0.35$, and we get $\xi=\{0.35, 0.52, 1.32\}$ for ``Undo Dancer", ``Shark",  and ``Hall", respectively.  It can be noticed that ``Undo Dancer" is the simplest sequence  for view synthesis (i.e., it has a small $\xi$ value) while ``Hall" is the most difficult one (i.e., it has a large $\xi$ value).    Fig.~4 depicts the achieved quality in view synthesis for both the experimental results and the theoretical model in \eqref{eq:distortion_DIBR}. 

Finally, in Table \ref{tab:video_character}, we summarize how much each video sequence is affected by both the coding and the synthesis process. For example, the ``Shark" video has few artifacts due to the compression. However the virtual synthesis is usually pretty challenging because of large dissimilarity among neighbouring camera views.

\begin{figure} 
\centering
\subfigure[Shark]{ \includegraphics[width=.45\linewidth,draft=false]{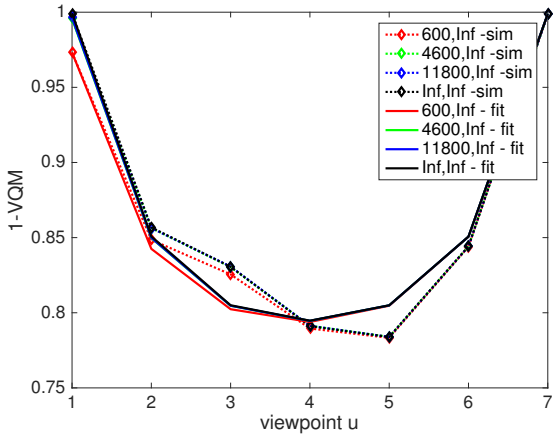}}
\hfill
\subfigure[Undo Dancer]{ \includegraphics[width=.45\linewidth,draft=false]{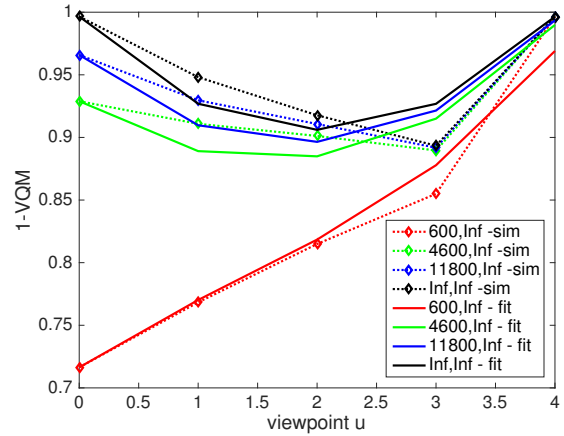}}
\hfill
\subfigure[Hall]{ \includegraphics[width=.45\linewidth,draft=false]{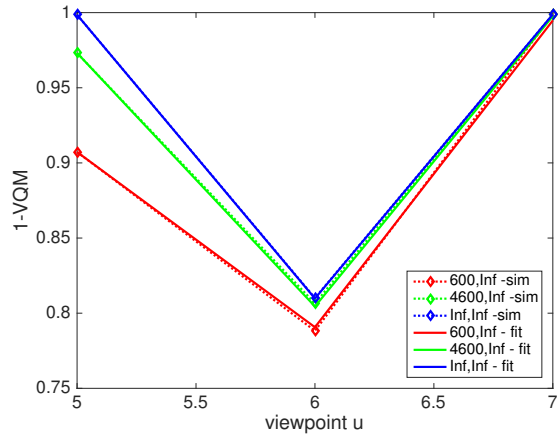}}
\hfill
\caption{Validation of the model for the virtual distortion for different video sequences and different quality of the anchor views. Dashed lines show  the simulation results, while solid lines show  the curve fitting model. The legend shows the encoding rate of left and right anchor views respectively, where an infinite rate means undistorted anchor view.}
\end{figure}
  
\begin{table} 
\begin{center}
\begin{tabular}{|c||c|c|c|c|c|}
  \hline
Video  		 &   Coding artifacts  	  & Synthesis artifacts  \\
  \hline   
 Shark 		&     low		&  medium \\  
 Undo Dancer 	&     high 	        &  low   \\
 Hall	 		&    medium 	& high  \\
  \hline
  \end{tabular}
\end{center}
\caption{Level of artifacts due to the coding as well as the synthesis processes.  }\label{tab:video_character}
\end{table} 

\clearpage

\section{Suboptimality under constraints C1 and C2.}
\label{sec:proofOpt}
 
We comment on the equivalence   between \eqref{eq:problem_form} and  \eqref{eq:problem_form_2} under the {\bf C1}  and  the {\bf C2} constraints described above.  The  {\bf C2} constraint  does not preclude optimality in problem   \eqref{eq:problem_form_2}. No optimal set of representations   leaves a viewpoint in the navigation window with no anchor views  in the original problem in \eqref{eq:problem_form}. This would lead to a zero satisfaction for the considered viewpoint.  
For the  {\bf C1} constraint, we consider the following illustrative example.  Let assume that   $v_1, v_2,$ and $v_3$ are camera views downloaded by the clients, then the {\bf C1}  precludes the scenarios  in which  viewpoints in $[v_1,v_2]$ are synthesized by $(v_1,v_2)$ while   viewpoints in $[v_2,v_3[$ are synthesized by  $(v_1,v_3)$.  If a representation is downloaded, it is used both as left or right reference view.  If anchor views have the same distortion, this condition still preserves the optimality of the problem   \eqref{eq:problem_form}. In the  case of different distortions among the downloaded representations,    the {\bf C1} constraint  still leads to optimality for most of the typical cases.   The main intuition is that if one camera view is selected only as right (or left) reference view, it is because there is a more far away camera that  is at a better distortion. However, as shown in     \cite{toni2015network}, this can happen if the other camera view is very close  or if the considered camera view is at very poor quality. None in the two cases   the camera view would actually be selected as best one. Therefore, the considered scenario would not happen in the optimal set. 
  
\section{Users Adaptation Logic}
\label{sec:adatLogic}
We now provide some further information about the adaptation logic implemented at the client side. We first notice that we consider users at regime (no startup or rebuffering phase); therefore each user makes a downloading request every temporal chunk. The user displaying the starting viewpoint $u$ at the downloading time will download the best set of MV navigation segments  $\mathcal{M}^*$ to reconstruct at her best the navigation window centered in $u$ given the bandwidth constraints imposed by the network connection. The best set to download $\mathcal{M}^*$ is derived by the following optimization problem
\begin{align}\label{eq:MVdistortion3}
 \mathcal{M}^* :  \  &\arg \min_{\substack{ \mathcal{M}:  l_L, l_R\subset \mathcal{T}   }} \sum_{m\in\mathcal{M}} D_{mc}(w)   \nonumber \\
&\text{s.t. } R_{\mathcal{M}} \leq B 
\end{align}

\begin{figure}[t]
\centering
\includegraphics[width=.75\linewidth,draft=false]{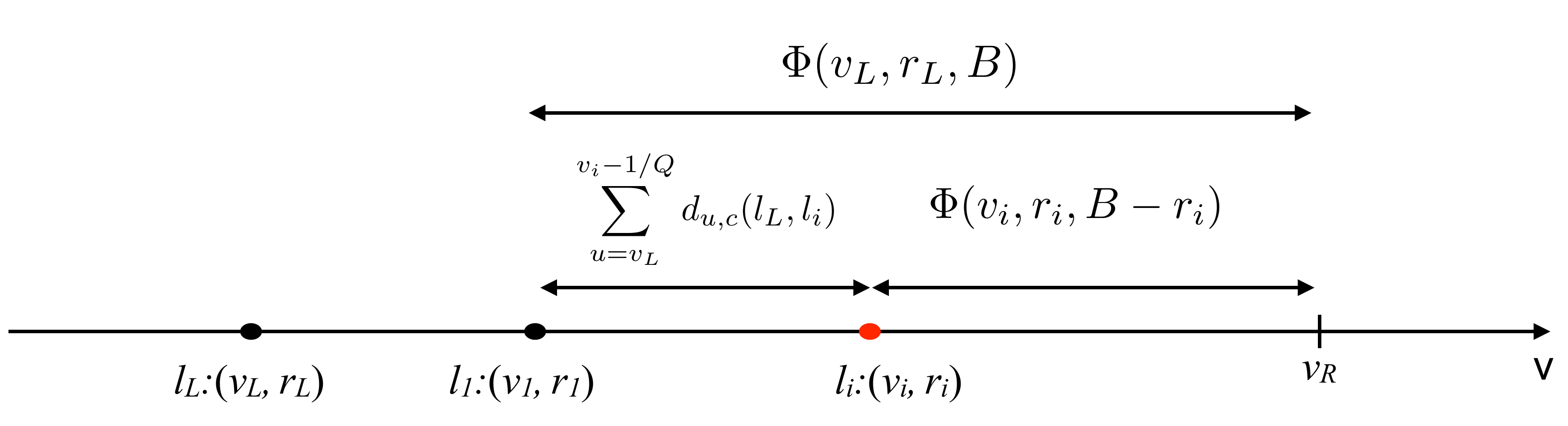} 
\caption{Visual illustration of the DP solving method.  }
\label{fig:DP}
\end{figure}

The above optimization problem can be solved in polynomial time with the following \ac{DP} algorithm.  We consider a user with a navigation window $w=[v_L,v_R]$. This NW can be ``cut" into segments given by the MV navigation segments dowloaded by the user. For example, in Fig. \ref{fig:DP}  downloading the MV navigation segment $m:\{l_L,l_1\}$ divides the region $[v_L,v_R]$ in two segments $[v_L,v_1[$ and $[v_1,v_R]$. If we then download another MV navigation segment $m_i:\{l_1,l_i\}$ we subdivide even more the region  $[v_1,v_R]$. Therefore, the optimization problem in \eqref{eq:MVdistortion3} can actually be solved by looking at the best ``cuts"  (or at the best view selection) in the range  $[v_L,v_R]$ with a maximum budget of $B$.  In \cite{toni2015network}, a DP solving method for this types of   view selection problems has been proposed. Here we adapt it to the case in which both view and rate need  to be optimized. 

We first define a recursive function $\Phi(v_L, r_L, b)$ as the minimum aggregate synthesized view distortion of views  between $[v_L,v_R[$  given that representation $l_L:(v_L,r_L)$ is already downloaded and the remaining bandwidth is $b$.  The algorithm selects  a new representation $l_i$ that subdivides the range of viewpoints in $[v_L,v_i[$ and $[v_i,v_R]$. Given the assumption of the MV navigation segment, i.e., given that any viewpoint between two consecutive representations are synthesized by these representations, the region $[v_L,v_i[$ is synthesized by $(l_L,l_i)$, leading to an aggregate distortion $  \sum_{u=v_L }^{v_i -1/Q} d_{u,c}(l_L,l_i)$ for the viewpoints in $[v_L,v_i[$. The remaining viewpoints in $[v_i,v_R[$ will be synthesized at the best distortion $ \Phi(v_i, r_i, b-r_i)$. This leads to the following recursion  
\begin{align}\label{eq:DP}
\Phi(v_L, r_L, b) =\min_{l_i:(v_i,r_i) \text{ s.t. } v_i > V_L} \left\{  \sum_{u=v_L }^{v_i -1/Q} d_{u,c}(l_L,l_i) +  \Phi(v_i, r_i, b-r_i)   \right\}\,.
\end{align}
The optimization problem in \eqref{eq:MVdistortion3} can therefore be solved with the following minimization
\begin{align}\label{eq:DP2}
\min_{ (v,r) \text{ s.t. }  v\leq v_L }\Phi(v, r, B-r)
\end{align}
that can be solved by DP  \cite{toni2015network}.

\end{document}